\documentclass[pre,twocolumn]{revtex4}
\usepackage{epsfig}
\usepackage{bm}
\usepackage{amsmath}

\newcommand{\bea}{\begin{eqnarray}}
\newcommand{\eea}{\end{eqnarray}}

\begin{document}

\title{Magneto-inertial range dominated by magnetic helicity in space plasmas}

\author{A. Bershadskii}

\affiliation{
ICAR, P.O. Box 31155, Jerusalem 91000, Israel
}

\begin{abstract}

 Magneto-inertial range dominated by magnetic helicity has been studied using results of numerical simulations, laboratory measurements, solar, solar wind, the Earth's and planets' magnetosphere observations (spacecraft measurements), and the global magnetic observatory network. The spectral data have been compared with the theoretical results based on the distributed chaos notion in the frames of the Kolmogorov-Iroshnikov phenomenology. The transition from magnetohydrodynamics to kinetics in the electron and Hall magnetohydrodynamics, and in a fully kinetic 3D approach, as well as in the solar wind, solar photosphere, and at the special events (reconnections, Kelvin-Helmholtz instability, isolated flux tube interchanges, etc.) in the magnetosphere of  Earth, Saturn, Jupiter, and Mercury has been studied using the above-mentioned data. Despite the considerable differences in the physical parameters and scales, the results of numerical simulations are in quantitative agreement with the observational data in the frames of the magneto-inertial range notion. Temporal variability of the magnetic field at Earth's surface under the influence of the ionosphere, magnetosphere, and solar wind has been also briefly discussed in this context.

\end{abstract}

\maketitle

\section{Introduction}

  There are three fundamental quadratic invariants of the ideal magnetohydrodynamics: total energy, magnetic, and cross helicity (see, for instance, Refs. \cite{mt},\cite{berg} and references therein). In this paper, we will concentrate on the former two (which also characterize the electron and Hall magnetohydrodynamics \cite{sch1},\cite{pm}). \\
  
  The magnetic helicity can be defined as 
\bea
 h_m &=& \langle {\bf a} {\bf b} \rangle  
\eea
where   ${\bf b} = [{\nabla \times \bf a}]$ is the fluctuating magnetic field, and $\langle ... \rangle$ denotes a spatial average (both ${\bf a}$ and ${\bf b}$ have zero means, and $\nabla \cdot {\bf a} =0$). 
 Under a uniform mean magnetic field ${\bf B_0}$ the magnetic helicity is not an invariant. In the paper Ref. \cite{wg} a modified magnetic helicity was introduced in the form
\bea
 \hat{h}_m &=& h_m + 2{\bf B_0}\cdot \langle {\bf A}  \rangle 
\eea
 where ${\bf B} = {\bf B_0} + {\bf b}$, ${\bf A} = {\bf A_0} +{\bf a}$, and ${\bf b} = [{\nabla \times \bf a}]$. It is shown in the paper \cite{wg} that at certain (rather weak) restrictions on the boundary conditions 
\bea
 \frac{d \hat{h}_m}{d t} &=&  0  
\eea
 in the ideal magnetohydrodynamics (see also Ref. \cite{shebalin}). \\

  It is understood now that deterministic chaos is a precursor of the randomized motions of fluids and plasmas. In Section II examples of the deterministic chaos in laboratory plasma,  the Hall magnetohydrodynamics (a direct numerical simulation), and solar wind will be provided (examples of deterministic chaos in the solar active regions will be provided in Section IX).\\ 
  
  In Section III distributed chaos that appears due to the natural randomization of the deterministic chaos under the domination of magnetic helicity will be studied using the results of the laboratory measurements, the direct numerical simulations (DNS), and solar observations.  \\
 
\begin{figure} \vspace{-0.5cm}\centering
\epsfig{width=.45\textwidth,file=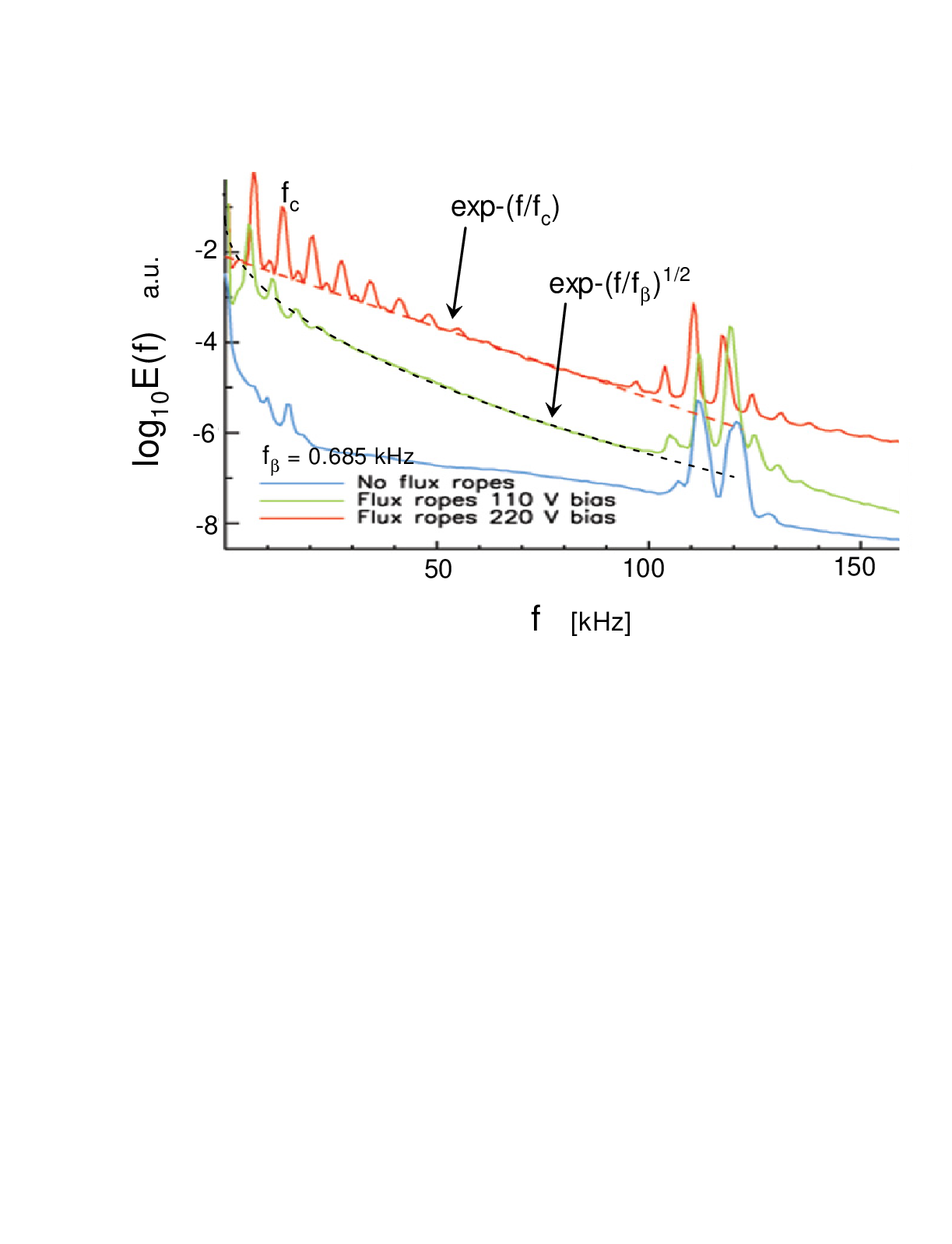} \vspace{-5.2cm}
\caption{Traces of average magnetic field power spectra measured in a laboratory experiment (the Large Plasma Device at UCLA) studying chaos in motions of magnetic flux ropes immersed in a magnetized helium plasma.} 
\end{figure}
 
   In hydrodynamic turbulence existence of an inertial range of scales with statistical characteristics depending only on the kinetic energy dissipation rate $\varepsilon$ (but not on the molecular viscosity) is expected for the high Reynolds numbers \cite{my}. Analogously, in magnetohydrodynamics (MHD) a magneto-inertial range of scales dominated by magnetic helicity can be introduced. In this case, we have two governing parameters the total energy (kinetic plus magnetic) dissipation rate $\varepsilon$ and the magnetic helicity dissipation rate $\varepsilon_h$. A similar situation (two governing parameters - kinetic dissipation rate and its analogy for the passive scalar) was considered for the so-called inertial-convective range of passive scalar turbulent mixing (the Corrsin-Obukhov approach \cite{my}). In Section IV we will use an analogous approach to the magneto-inertial range of scales, but in the frames of the distributed chaos notion and the Kolmogorov-Iroshnikov phenomenology. 
   
   In Section V the theoretical results will be compared with the results of the laboratory measurements and the direct numerical simulations with the electron and Hall magnetohydrodynamics, and a fully kinetic 3D model. 
   
   In Section VI the theoretical results will be compared with the results of observations in the solar wind (the spacecraft measurements produced at different distances from the Sun: from 0.16 AU to 4.5 AU). 
   
    In Sections VII and VIII the smooth transition from large-scale pure magnetohydrodynamics to kinetics will be studied in the Earth’s and planets' magnetosphere respectively (in particular at the special events: reconnections, Kelvin-Helmholtz vortices, and isolated flux tube interchanges, etc).
 
  In Section IX the theoretical results will be compared with the results of observations in the solar active regions.\\
  
    In a wide range of time scales temporal variability of the magnetic field at Earth's surface is dominated by the variability of the external magnetic field arising from the interaction of the near-Earth solar wind with the Earth’s magnetosphere and its inner edge - ionosphere. In Section X this variability will be studied using the magneto-inertial range of scales notion and the data obtained with the global magnetic observatory network. 
  
  In Section XI conclusions summarizing the results obtained in the previous sections will be presented.

\section{Deterministic chaos in the magnetized plasmas}

\begin{figure} \vspace{-0.8cm}\centering
\epsfig{width=.46\textwidth,file=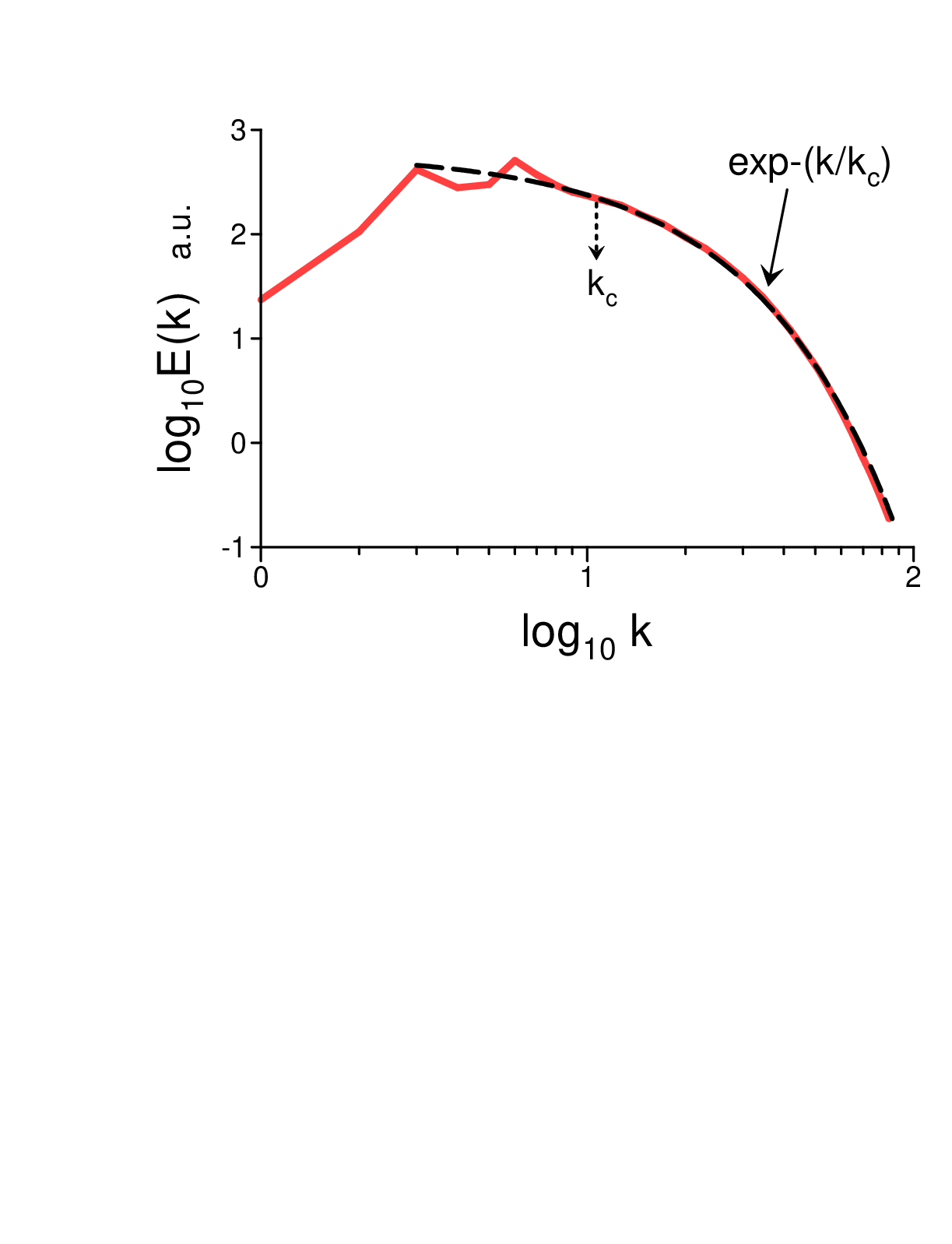} \vspace{-5.2cm}
\caption{ Magnetic energy spectrum at the saturation stage of a small-scale Hall MHD dynamo.} 
\end{figure}

   Deterministic chaos (at the onset of turbulence) is often characterized by the exponential power spectra (both temporal and spatial)\cite{fm}-\cite{kds}
\bea
 E(f) \propto \exp-(f/f_c) 
\eea
 and/or 
\bea
E(k) \propto \exp-(k/k_c)  
\eea
where $f$ is frequency, $k$ is wavenumber. \\

   Figure 1, for instance, shows (in the linear-log scales) the traces of average magnetic field power spectra measured in a laboratory experiment (the Large Plasma Device at UCLA) studying chaos in motions of magnetic flux ropes (magnetic structures with helical currents and magnetic fields) immersed in a magnetized helium plasma \cite{002}. The magnetic flux ropes rotate about a central axis in a cylindrical vessel, writhe about each other, twist about themselves, and are kink unstable. Collisions between the magnetic ropes result in magnetic field lines reconnection. \\
   
   According to Taylor's ``frozen in'' hypothesis, the experimental spectrum shown in Fig. 1 should be interpreted as a wavenumber spectrum with $k \simeq 2\pi f/V_0$, where $V_0$ is the mean speed of the spatial magnetic structures passing the probe (cf., for instance, Refs. \cite{mm},\cite{sbl} and references therein). The dashed straight line is drawn through the upper curve in Fig. 1 to indicate the exponential spectrum Eq. (5) corresponding to the spatial {\it deterministic} chaos (the stretched exponential spectrum for the flux ropes at 110 V will be explained in the next Section).\\

  In a recent paper Ref. \cite{hal} the authors numerically simulated (DNS) a small-scale magnetohydrodynamic dynamo using the three-dimensional `incompressible' Hall equations in the form:
\bea
\partial_t {\bf u} + ({\bf u}\cdot \nabla) {\bf u}  \!\!&=&\!\!  -\nabla P + \left({\bf b}\cdot \nabla\right){\bf b} +\nu\nabla^2 {\bf u} + {\bf f} \\
\partial_t {\bf b} + \left({\bf u}\cdot {\bf \nabla}\right){\bf b} \!\!&=&\!\! \left({\bf b}\cdot \nabla\right){\bf u} + \eta\nabla^2 \bf{b} \nonumber \\
 && \!\!\!\!\!\!\!\!\! \!\!\!\!\!\!\!\!\!\!\!\!\!\!\!\!\!\!\!   + d_i\left({\bf j}\cdot {\bf \nabla}\right){\bf b} -d_i\left({\bf b}\cdot \nabla\right){\bf j} \\
\nabla \cdot {\bf u}  &=& 0, ~~ {\bf \nabla} \cdot {\bf b}  = 0 
\eea

where ${\bf u}$ is the velocity field, $P = \left(p +\frac{{\bf b}^2}{2}\right)$, $p$ is the kinetic pressure (normalized by the constant plasma density $\rho$), ${\bf f}$ is the hydrodynamic large-scale Taylor-Green external forcing:  ${\bf f} \equiv f_0 [ \sin(k_f x)\cos(k_f y)\cos(k_f z) {\hat {\bf x}} - \cos(k_fx)\sin(k_fy)\cos(k_fz) {\hat {\bf y}}]$ (here $k_f = 2$), $\eta$ and $\nu$ are the magnetic diffusivity and kinematic viscosity correspondingly, ${\bf b}= {\bf B}/\sqrt{\mu_0 \rho}$ is the normalized magnetic field, ${\bf j}= \nabla \times {\bf b}$, and $d_i$ is the ion inertial length. \\

   The last two terms on the right-hand side of Eq. (7 ) represent the Hall effect. In the ideal Hall MHD ions are not already tied to the magnetic field due to the ions' inertia, while the electrons still are. The ion inertial length $d_i$ corresponds to the spatial scale at which electrons and ions decouple.\\

  Periodic boundary conditions have been used in the DNS for all spatial directions and a random seed magnetic field was introduced in a statistically stationary chaotic/turbulent motion of the fluid at small scales to initiate the small-scale magnetohydrodynamic dynamo. Magnetic Prandtl number $P_m =1$. 

   Figure 2 shows the magnetic energy spectrum at the saturation stage of the Hall MHD dynamo. The spectral data were taken from Fig. 3 of the Ref. \cite{hal}. The dashed curve indicates the exponential spectrum Eq. (5) (deterministic chaos). The dotted arrow indicates the position of the characteristic scale $k_c$. \\

In paper Ref. \cite{pg} the transition from magnetohydrodynamics (large spatial scales) to kinetics (small spatial scales) in the solar wind was studied with emphasis on the effects related to the magnetic helicity (see also next Section). \\
   
   Figure 3 shows the trace magnetic power spectrum measured at the Ulysses Northern Polar Pass ($R = 2.36$ AU) in a fast (high-speed) solar wind. The spectral data were taken from Fig. 2 of the Ref. \cite{pg}.  The dashed vertical line indicates the frequency position (using Taylor's ``frozen-in'' hypothesis) of the thermal ion gyroradius $\rho_i$. The dotted straight line in the log-log scales indicates the Kolmogorov-like power law $E(f) \propto f^{-5/3}$ (or according to Taylor's ``frozen-in'' hypothesis $E(k) \propto k^{-5/3}$) for the large scales. The short-dash curve indicates correspondence to the exponential spectrum Eq. (5), i.e. to the deterministic chaos. The position of the characteristic frequency of the deterministic chaos $f_c$ is indicated by the dotted arrow (or according to Taylor's ``frozen-in'' hypothesis the position of the characteristic wavenumber $k_c$) (cf Fig. 2).\\ 

  One can also see examples of the deterministic chaos in the magnetic field at the onset of emerging solar active regions in Figs. 37 and 38 (Section IX).
  
\begin{figure} \vspace{-1.3cm}\centering \hspace{-1cm}
\epsfig{width=.52\textwidth,file=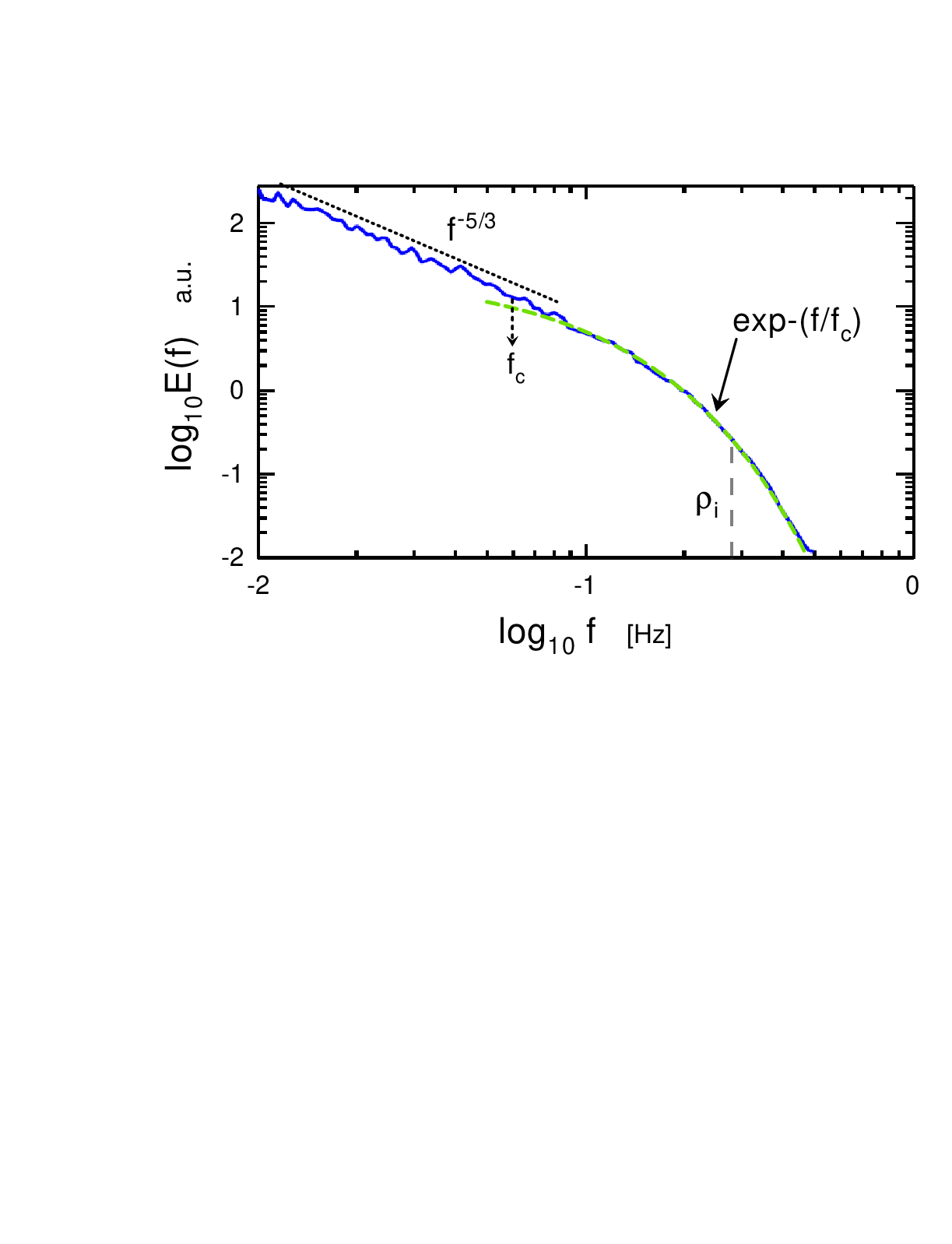} \vspace{-5.8cm}
\caption{Trace magnetic power spectrum measured at the Ulysses Northern Polar Pass ($R = 2.36$ AU). }
\end{figure}
  
\section{Distributed chaos dominated by magnetic helicity}   
   
     At a sufficiently strong change of the parameters driving the magnetic field fluctuations dynamics the characteristic scale $k_c$ in Eq. (5) can become randomly fluctuating. Then to compute the magnetic power spectra one needs an ensemble averaging
\bea
E(k) \propto \int_0^{\infty} P(k_c) \exp -(k/k_c)dk_c 
\eea 
where a probability distribution $P(k_c)$ characterizes the fluctuations of the characteristic scale $k_c$. 

To find the probability distribution $P(k_c)$ in the plasma dominated by the magnetic helicity one can use a scaling relationship between characteristic values of the magnetic field $B_c$ and $k_c$
\bea
B_c \propto |h_m|^{1/2} k_c^{1/2}   
\eea
obtained with the dimensional considerations. If the $B_c$ has a normal distribution (Gaussian with zero mean \cite{my}), then $k_c$ has the chi-squared ($\chi^{2}$) distribution
\bea
P(k_c) \propto k_c^{-1/2} \exp-(k_c/4k_{\beta}) 
\eea
where $k_{\beta}$ is a new characteristic constant. 

   Substituting Eq. (11) into Eq. (9) one obtains
\bea
E(k) \propto \exp-(k/k_{\beta})^{1/2}  
\eea
This is the magnetic energy spectrum for a {\it distributed} chaos in the presumably non-dissipative range of scales dominated by the magnetic helicity. \\

  In a special interesting case with global reflectional symmetry, the global (mean) magnetic helicity is identically equal to zero (or negligible). The point-wise magnetic helicity can be not identically equal to zero even in this case. Namely, the spontaneous breaking of the local reflectional symmetry (and related spontaneously generated local helical fluctuations) is an inherent property of the strongly chaotic/turbulent flows. The appearance of the blobs having non-zero blob's kinetic/magnetic helicity can accompany this process (see, for instance, Refs. \cite{mt},\cite{kerr}-\cite{ber1} and references therein). Some of the magnetic blobs are bounded by the so-called magnetic surface where ${\bf b_n}\cdot {\bf n}=0$ (${\bf n}$ is a unit normal to the blob's boundary). Sign-defined magnetic helicity of such a blob is an ideal invariant \cite{mt}. Let us number the blobs and denote their helicity as $H_j^{\pm}$ depending on their sign ($j$ is the number of a blob):
\bea
 H_j^{\pm} &=& \int_{V_j} ({\bf a} ({\bf x},t) \cdot  {\bf b} ({\bf x},t)) ~ d{\bf x} 
\eea 
  
  Then let us consider the ideal adiabatic invariant 
\bea
{\rm I^{\pm}} &=& \lim_{V \rightarrow  \infty} \frac{1}{V} \sum_j [H_{j}^{\pm}]  
\eea 
where the summation in Eq. (14) is over the blobs with a certain sign only (`+' or `-') and $V$ is the total spatial volume of the blobs considered for the sum.  \\

Due to the global reflectional symmetry ${\rm I^{+}} \simeq - {\rm I^{-}}$. For the case of global reflectional symmetry the adiabatic invariant ${\rm I^{\pm}}$ Eq. (14) 
can replace  $h_m$ in Eq. (10)
\bea
B_c \propto |{\rm I^{\pm}}|^{1/2} k_c^{1/2}   
\eea

This replacement results in the same stretched exponential spectrum Eq. (12) for this case.\\

\begin{figure} \vspace{-1.2cm}\centering
\epsfig{width=.46\textwidth,file=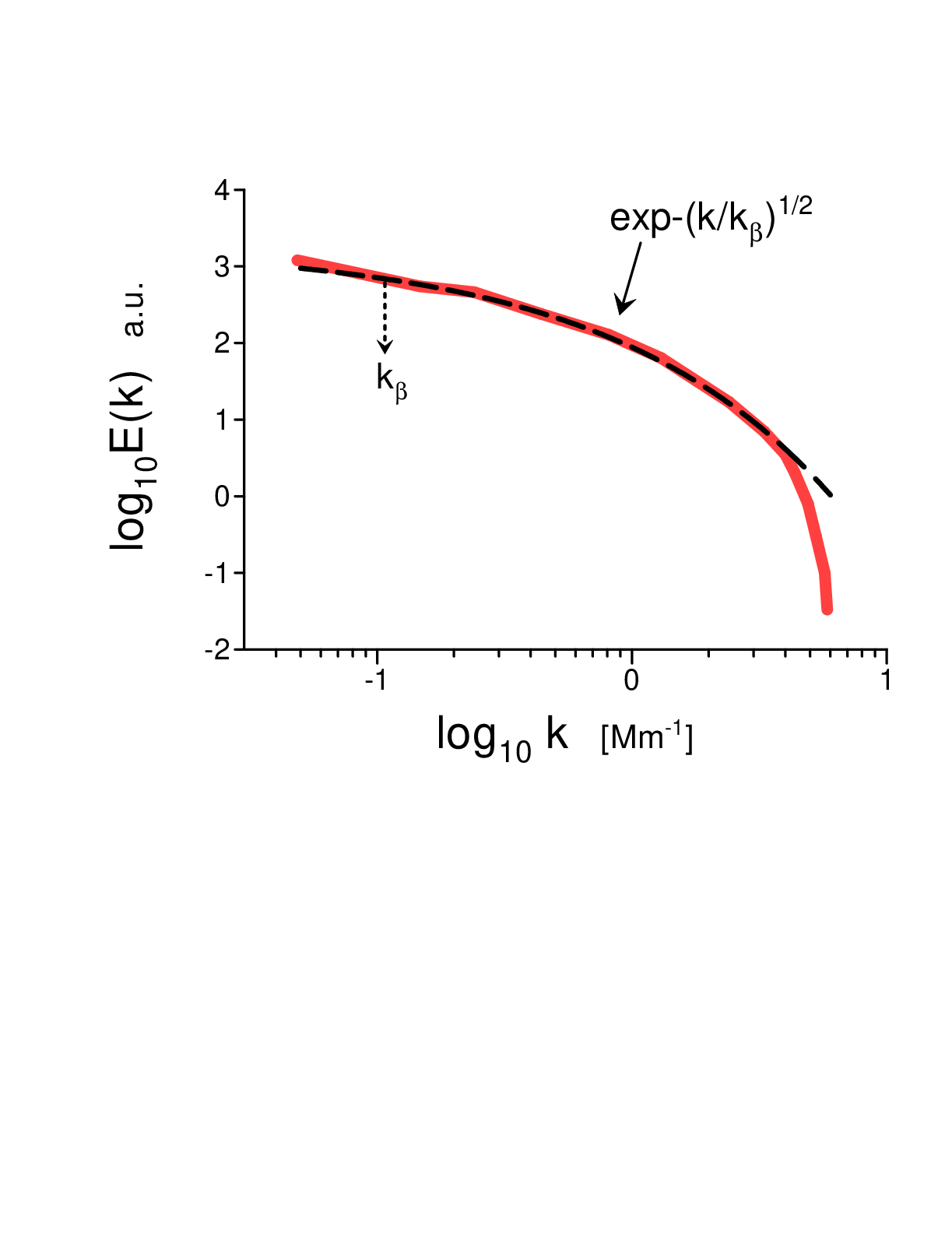} \vspace{-4.3cm}
\caption{The spot and noise corrected magnetic power spectrum for the solar active region NOAA 8375.} 
\end{figure}
\begin{figure} \vspace{-0.5cm}\centering
\epsfig{width=.46\textwidth,file=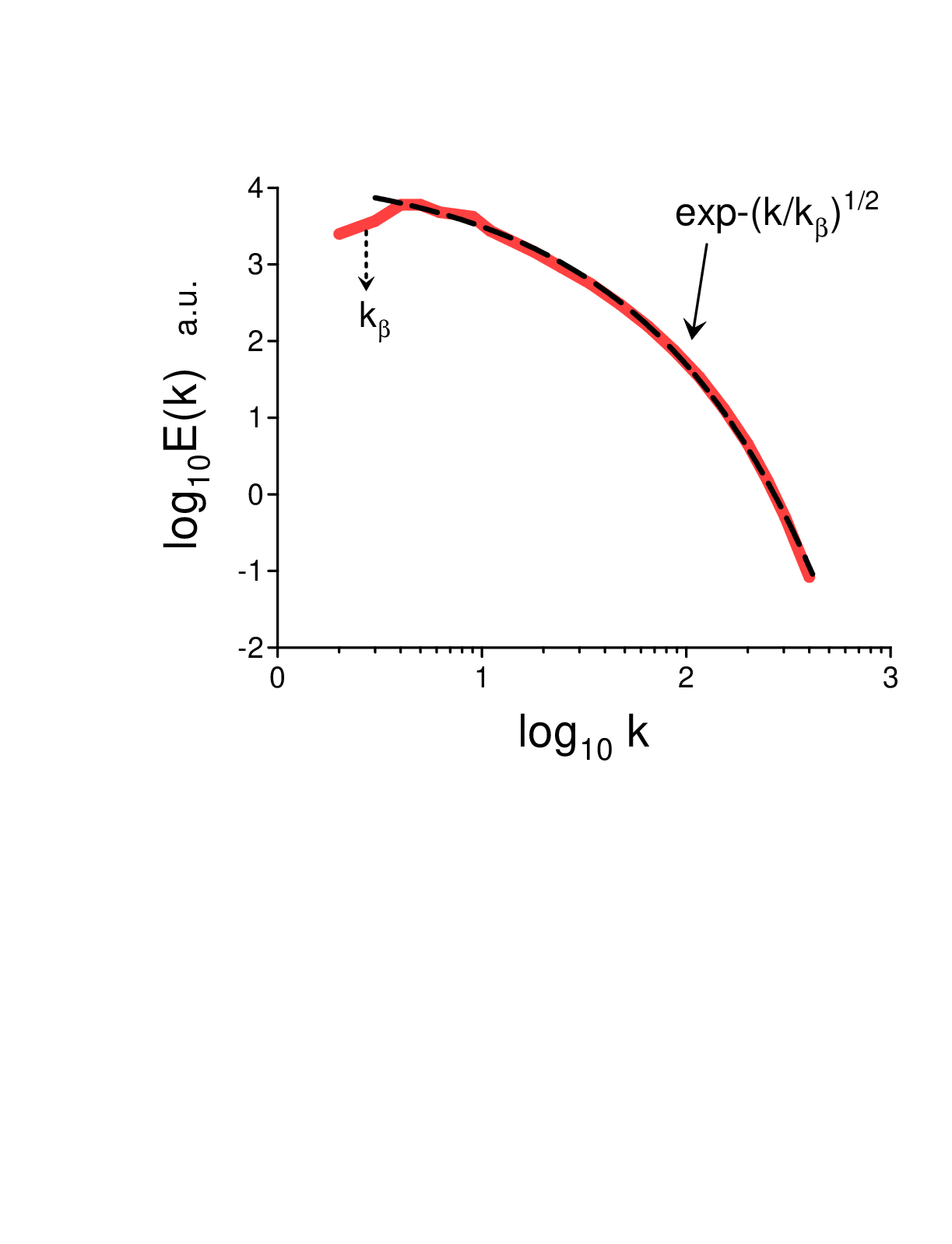} \vspace{-4.2cm}
\caption{The DNS computed magnetic energy spectra for forced incompressible isotropic homogeneous MHD for $Re_{\lambda} =Re_{\lambda}^m=909$.  } 
\end{figure}
\begin{figure} \vspace{-0.85cm}\centering
\epsfig{width=.45\textwidth,file=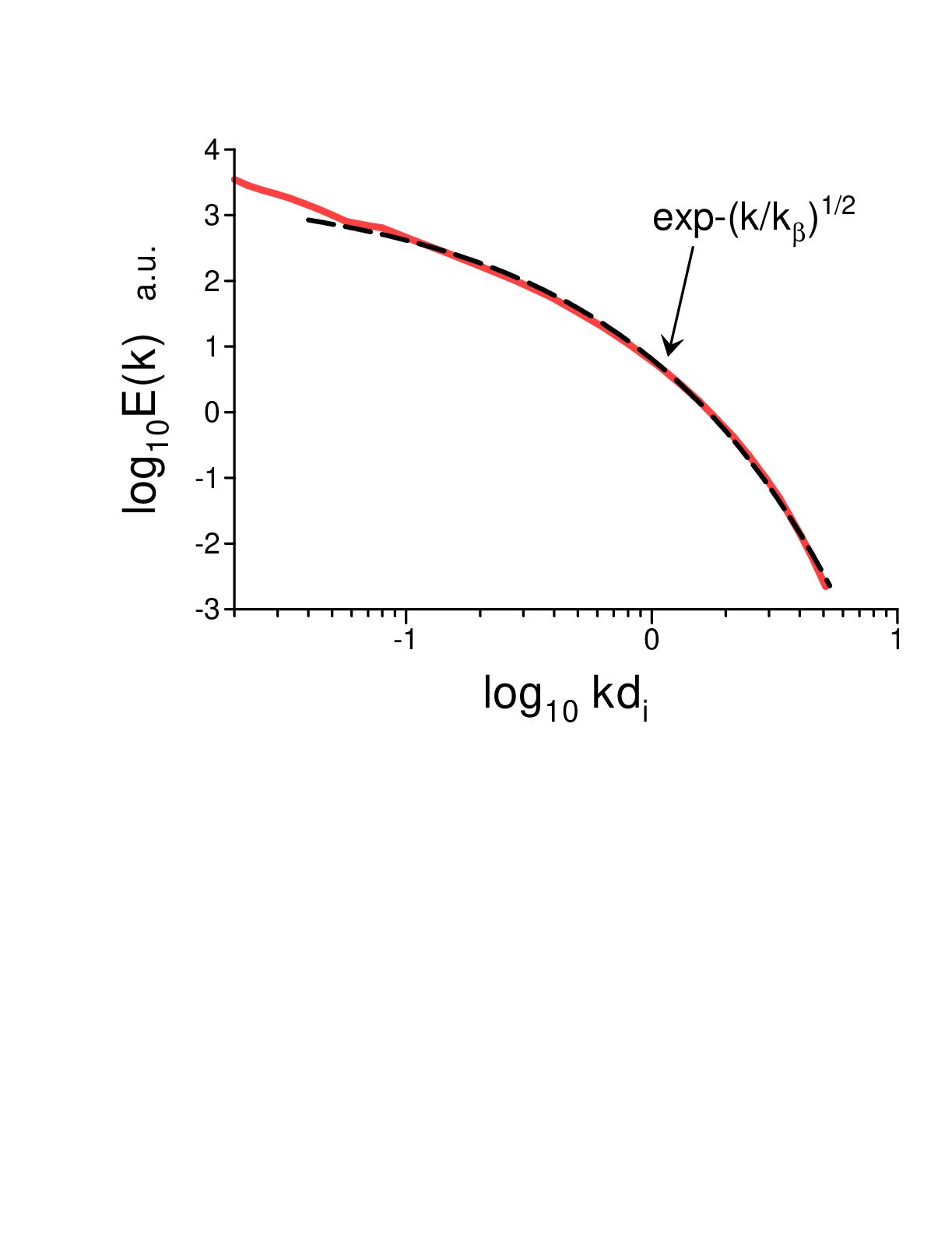} \vspace{-4.45cm}
\caption{The DNS computed magnetic energy spectra for decaying compressible weakly anisotropic Hall MHD at the time of maximal dissipation. } 
\end{figure}

  The dashed curve is drawn through the middle curve in Fig. 1 to indicate the stretched exponential spectrum Eq. (12) corresponding to the spatially distributed chaos dominated by magnetic helicity. Since the experimental spatial domain is not enclosed completely by a magnetic flux surface the ${\bf a}$ is not gauge invariant. In this case, a gauge-invariant relative magnetic helicity with an appropriate reference magnetic field (the axial background magnetic field - constant in time and space outside the experimental vessel,) can be used instead \cite{bf},\cite{jc},\cite{dg}.\\
 
  It is well known that multiple flux ropes extend from the solar surface into the solar photosphere and corona. Coronal Mass Ejections can be triggered by the multiple flux ropes collisions \cite{amar}. \\
  
    In Ref. \cite{abr} results of the space-based (SOHO/MDI) spectral measurements of the longitudinal magnetic fields in the solar active region NOAA 8375 (made on 4 November 1998) were reported. At the time of the measurements, the active region was located at the center of the solar disk. To avoid the strong contribution of the sunspot to the power spectrum the authors of the Ref. \cite{abr} blocked out the sunspot. Figure 4 shows the spot and noise-corrected magnetic power spectrum (the spectral data were taken from Fig. 6b of the Ref. \cite{abr}). The dashed curve is drawn to indicate the stretched exponential spectrum Eq. (12) corresponding to the spatially distributed chaos dominated by magnetic helicity  (cf Ref. \cite{ber2}). \\
    
    Magnetic ropes are routinely ejected from the Sun. These magnetic flux ropes can transport a large amount of magnetic helicity into the solar wind (see Section VI). \\    

    Let us start a comparison of the laboratory and solar data with the results of DNS of isotropic homogeneous pure magnetohydrodynamics (Eqs. (6-8) without the Hall terms) reported in a paper Ref. \cite{ae}. The statistically stationary state with high $Re_{\lambda} =Re_{\lambda}^m=909$ was reached using the Taylor-Green forcing in a spatially periodic box. \\
    
    Figure 5 shows the magnetic energy spectrum computed in this DNS (the spectral data were taken from Fig. 1 of the Ref. \cite{ae}). The dashed curve in Fig. 5 indicates the stretched exponential spectrum Eq. (12) corresponding to the spatially distributed chaos dominated by magnetic helicity (cf Figs. 1,4 and 12).\\
    
   Figure 6 shows the magnetic energy spectrum computed in a DNS reported in a recent paper \cite{fer} (the spectral data were taken from Fig. 6a of the Ref. \cite{fer}). In this DNS a freely decaying compressible isothermal case was studied in the frames of the Hall MHD with a weak background magnetic field. The DNS was performed in a cubic spatial box with periodic boundary conditions and with initial conditions consisting of a superposition of random phase harmonic modes. The energy of these modes in Fourier space was localized in a sphere $1 < k < 3$. The initial Mach number $M_s = 0.25$ and $d_i = 0.02L_0$ (where $L_0 =2\pi$ is the size of the DNS cubic box). The spectrum was computed in the vicinity of the time when a maximal dissipation was reached. 

The dashed curve in Fig. 6 indicates the stretched exponential spectrum Eq. (12) corresponding to the spatially distributed chaos dominated by magnetic helicity (cf Figs. 1,4). \\

 \section{Magneto-inertial range of scales}

   Conditions in the solar wind are significantly different from those in the solar atmosphere. The main difference is the high speed of the plasma moving away from the Sun. Therefore, a different phenomenology - magneto-inertial range (see Introduction) should be used for the solar wind. In this case, the estimate Eq. (10) should be replaced by 
\bea
 B_c \propto \varepsilon_{h_m}^{1/2} ~\varepsilon^{-1/6}~k_c^{1/6}  
\eea
 (cf the Corrsin-Obukhov phenomenology for the inertial-convective range \cite{my}).  \\

 If a significant mean magnetic field is present, then the energy dissipation rate $\varepsilon$ in Eq. (16) should be replaced by a more appropriate parameter $(\varepsilon \widetilde{B}_0)$ (here  $\widetilde{B}_0 = B_0/\sqrt{\mu_0\rho}$ is the normalized mean magnetic field, having the same dimension as velocity ) \cite{ir}. The dimensional considerations, in this case, result in
\bea
 B_c \propto \varepsilon_{\hat{h}_m}^{1/2}~ (\varepsilon \widetilde{B}_0)^{-1/8}  k_c^{1/8} 
\eea
 with the modified magnetic helicity (see the Eq. (2)) dissipation rate $\varepsilon_{\hat{h}_m}$. However, the dimensional estimation Eq. (16) can be still valid for some cases even with a strong external magnetic field.\\
 
 The estimates Eq. (16-17) can be generalized as
\bea
 B_c \propto k_c^{\alpha}   
 \eea
  
   Let us, following Eq. (12), seek the spectrum for the distributed chaos as a stretched exponential one
\bea
E(k) \propto \int_0^{\infty} P(k_c) \exp -(k/k_c)dk_c \propto \exp-(k/k_{\beta})^{\beta} 
\eea

  For large $k_c$ the probability distribution $P(k_c)$ can be estimated from the Eq. (19) as \cite{jon}
\bea
P(k_c) \propto k_c^{-1 + \beta/[2(1-\beta)]}~\exp(-\gamma k_c^{\beta/(1-\beta)}) 
\eea    
  
   Then, using some algebra, the relationship between the exponents $\beta$ and $\alpha$ can be readily obtained from the Eqs. (18) and (20) for the normally distributed $B_c$
\bea
\beta &=& \frac{2\alpha}{1+2\alpha}  
\eea

  Hence, for $\alpha =1/6$ (see Eq. (16))
\bea
 E(k) \propto \exp-(k/k_{\beta})^{1/4}  
\eea
 and for $\alpha =1/8$ (see Eq. (17))
\bea
 E(k) \propto \exp-(k/k_{\beta})^{1/5}  
\eea
  
  These are spatial magnetic power spectra in the magneto-inertial range of scales dominated by magnetic helicity.\\ 

\begin{figure} \vspace{-0.8cm}\centering
\epsfig{width=.47\textwidth,file=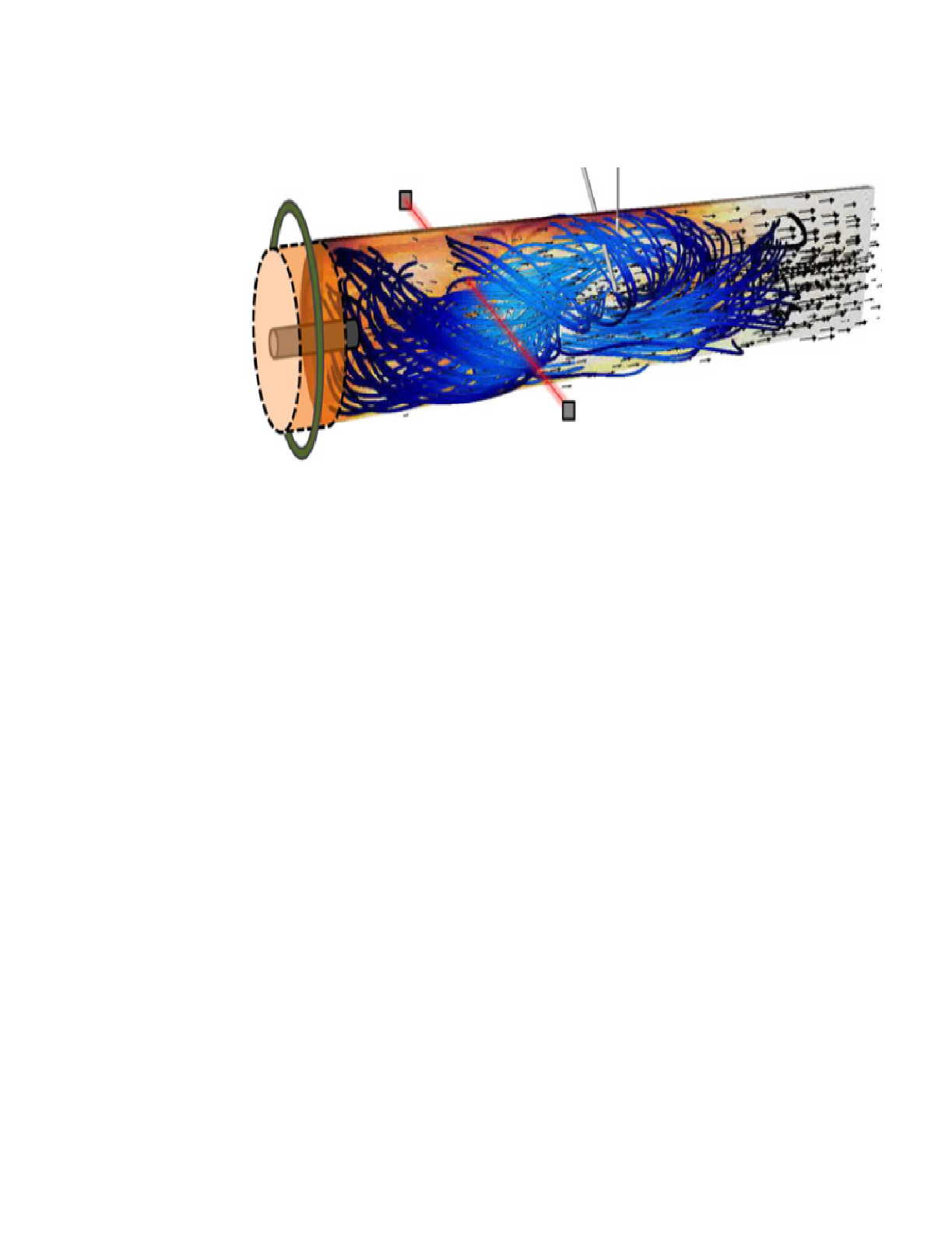} \vspace{-6.2cm}
\caption{A sketch of the MHD (plasma gun) wind tunnel.} 
\end{figure}
\begin{figure} \vspace{-0.5cm}\centering
\epsfig{width=.45\textwidth,file=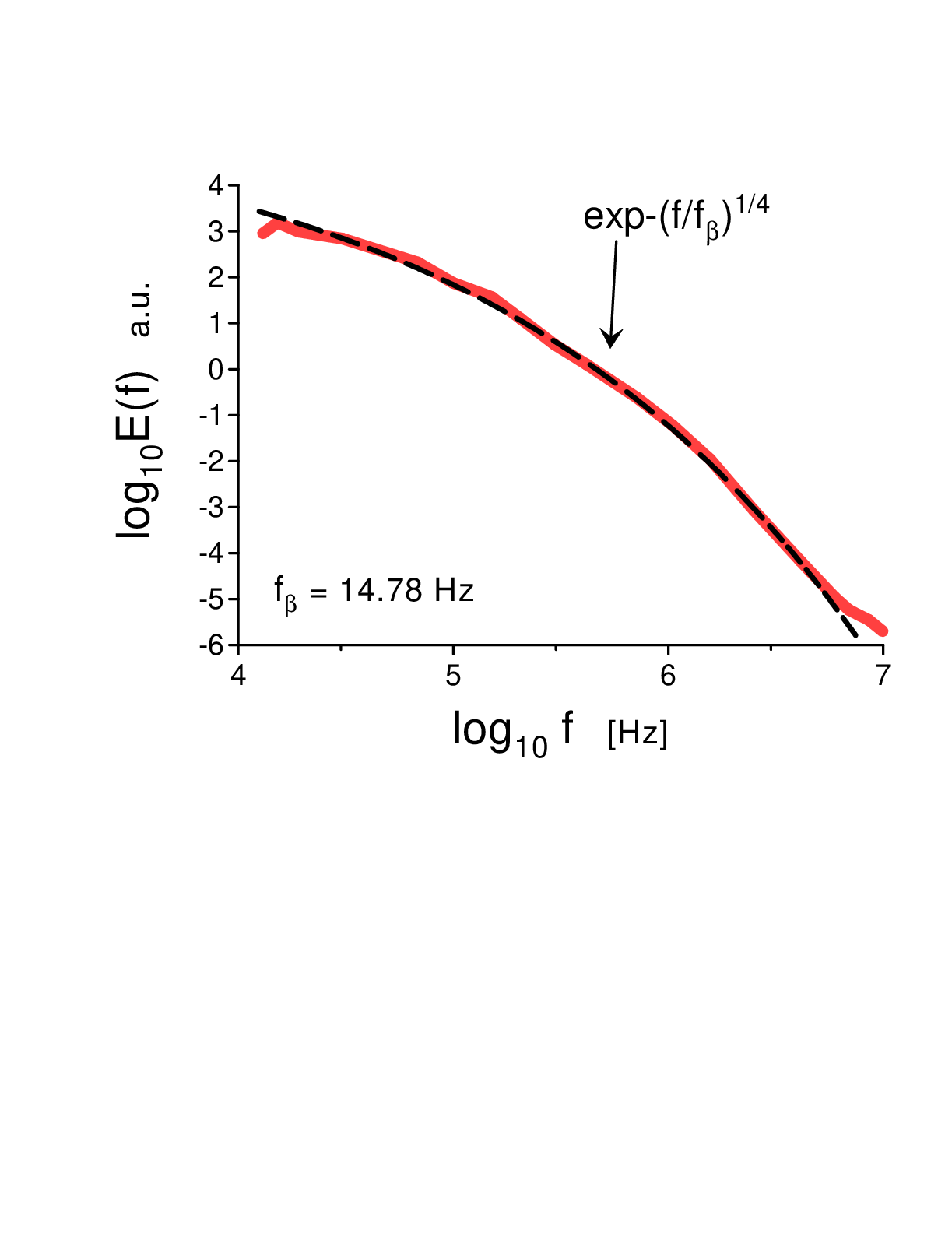} \vspace{-4.2cm}
\caption{Ensemble averaged total magnetic energy spectrum measured in the MHD (plasma gun) wind tunnel experiment.} 
\end{figure}
\begin{figure} \vspace{-1cm}\centering\hspace{-1cm}
\epsfig{width=.49\textwidth,file=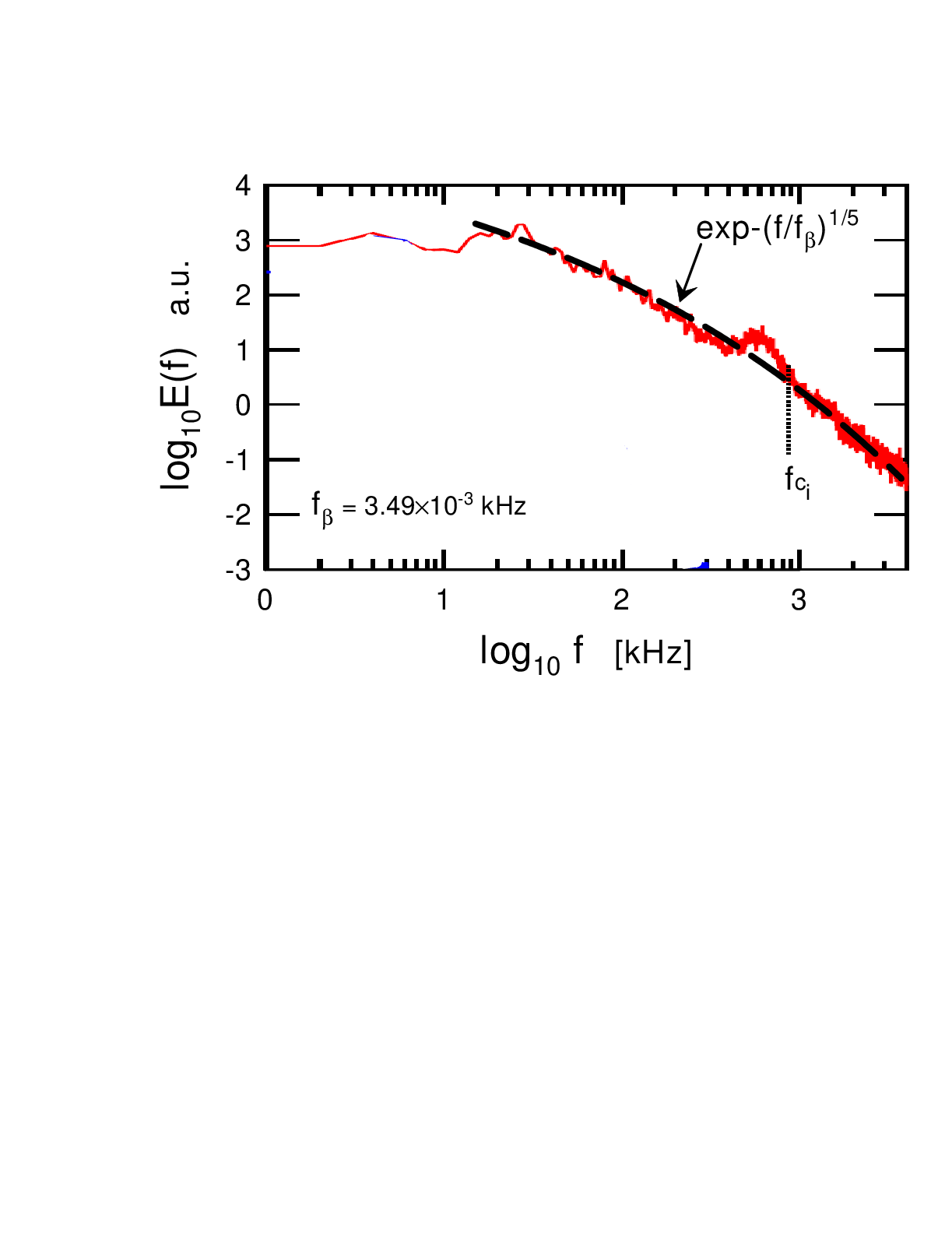} \vspace{-5.3cm}
\caption{Power spectrum of the vertical component of the magnetic field measured within the plasma edge. }
\end{figure}
  
 \section{Laboratory measurements vs DNS}
 
 \subsection{Laboratory measurements}
 
Figure 7 (adapted from the Ref. \cite{sbl}) shows a sketch of the MHD (plasma gun) wind tunnel. This experiment can imitate situations similar to those typical for solar wind. The blue lines in the figure show twisting of the magnetic field lines due to magnetic helicity conservation.\\

  Figure 8 shows the ensemble-averaged total magnetic energy spectrum measured in the experiment. The spectral data were taken from Fig. 15a of the Ref. \cite{sbl}. Taylor's ``frozen in'' hypothesis can be used to transform the measured frequency spectrum to the more adequate in this case wavenumber spectrum. The dashed curve indicates correspondence to the stretched exponential spectrum Eq. (22).\\
  
    In recent Ref. \cite{rich} results of an experiment with local magnetic helicity injection into a toroidal system (a spherical tokamak startup) have been reported and the physical mechanisms converting the injected magnetic helicity into a large-scale current drive are experimentally studied. In this experiment, the intense electron current sources inject the magnetic helicity at the edge of a tokamak. Magnetic reconnections associated with the helicity-conserving instabilities convert the topology of the open field lines into a tokamak-like toroidal one. \\
 
  Figure 9 shows the power spectrum of the vertical component of the magnetic field measured within the plasma edge. The spectral data were taken from Fig. 1 of the Ref. \cite{rich}. The position of the local ion cyclotron frequency $f{\tiny c}_i$ is indicated by the dotted vertical line. The dashed curve indicates the stretched exponential spectrum Eq. (23).
  
 \subsection{MHD direct numerical simulations}  
 
\begin{figure} \vspace{-1.4cm}\centering
\epsfig{width=.45\textwidth,file=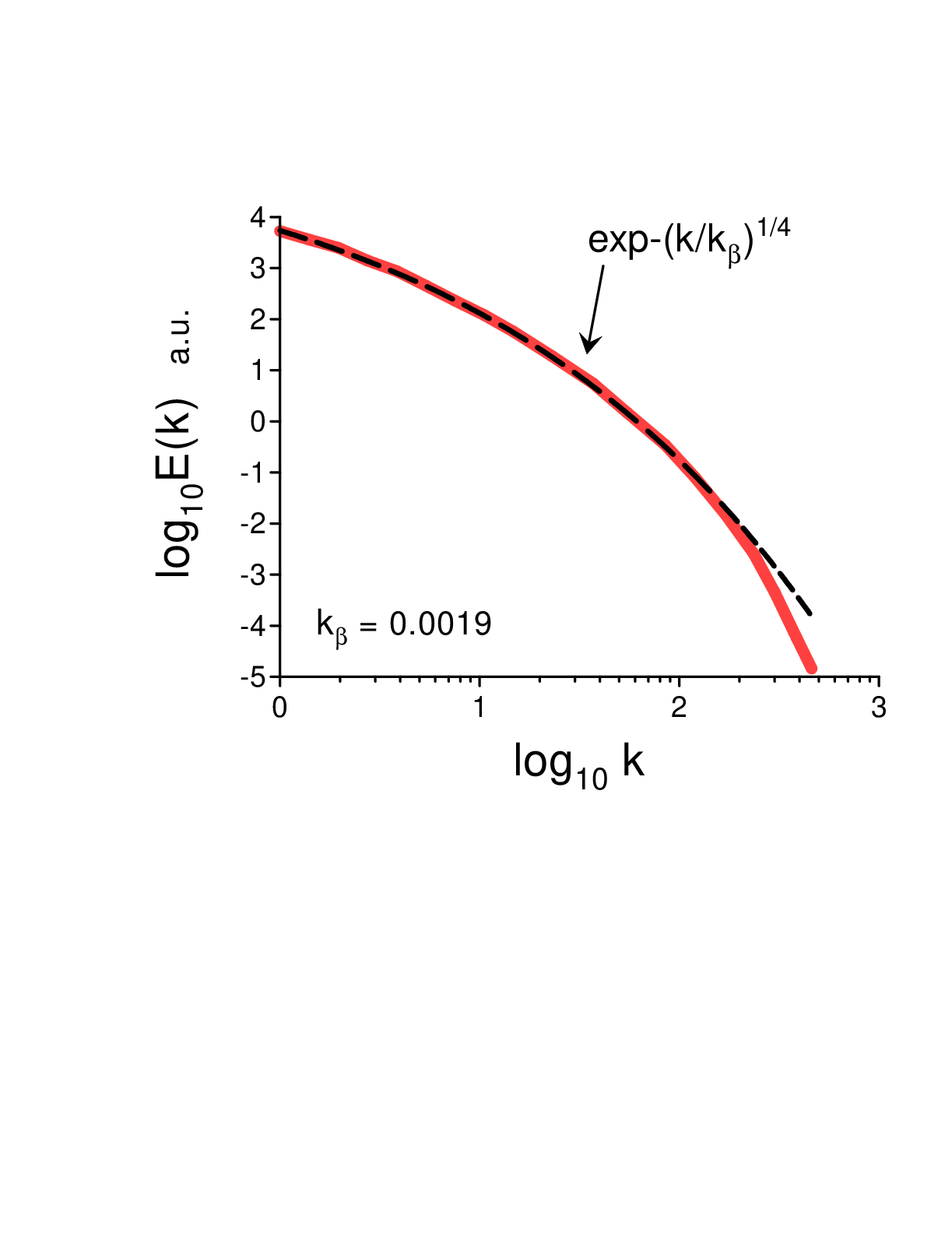} \vspace{-3.95cm}
\caption{Magnetic energy spectra for the freely decaying MHD DNS with considerable mean magnetic helicity.  }
\end{figure}
\begin{figure} \vspace{-0.5cm}\centering
\epsfig{width=.45\textwidth,file=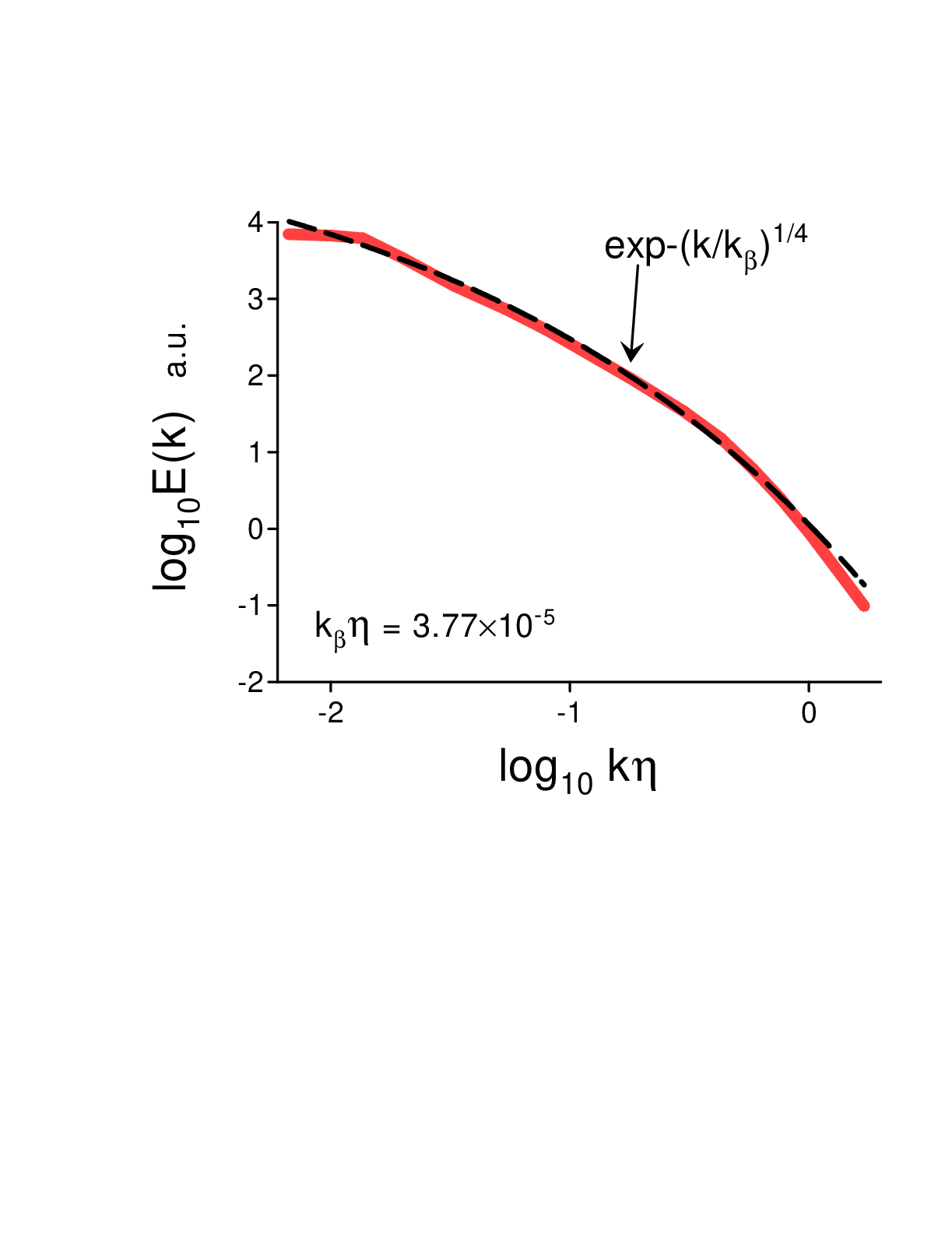} \vspace{-3.9cm}
\caption{Magnetic energy spectrum for the forced MHD DNS with considerable mean magnetic helicity (here $\eta$ is the dissipation scale).} 
\end{figure}
\begin{figure} \vspace{-0.5cm}\centering
\epsfig{width=.44\textwidth,file=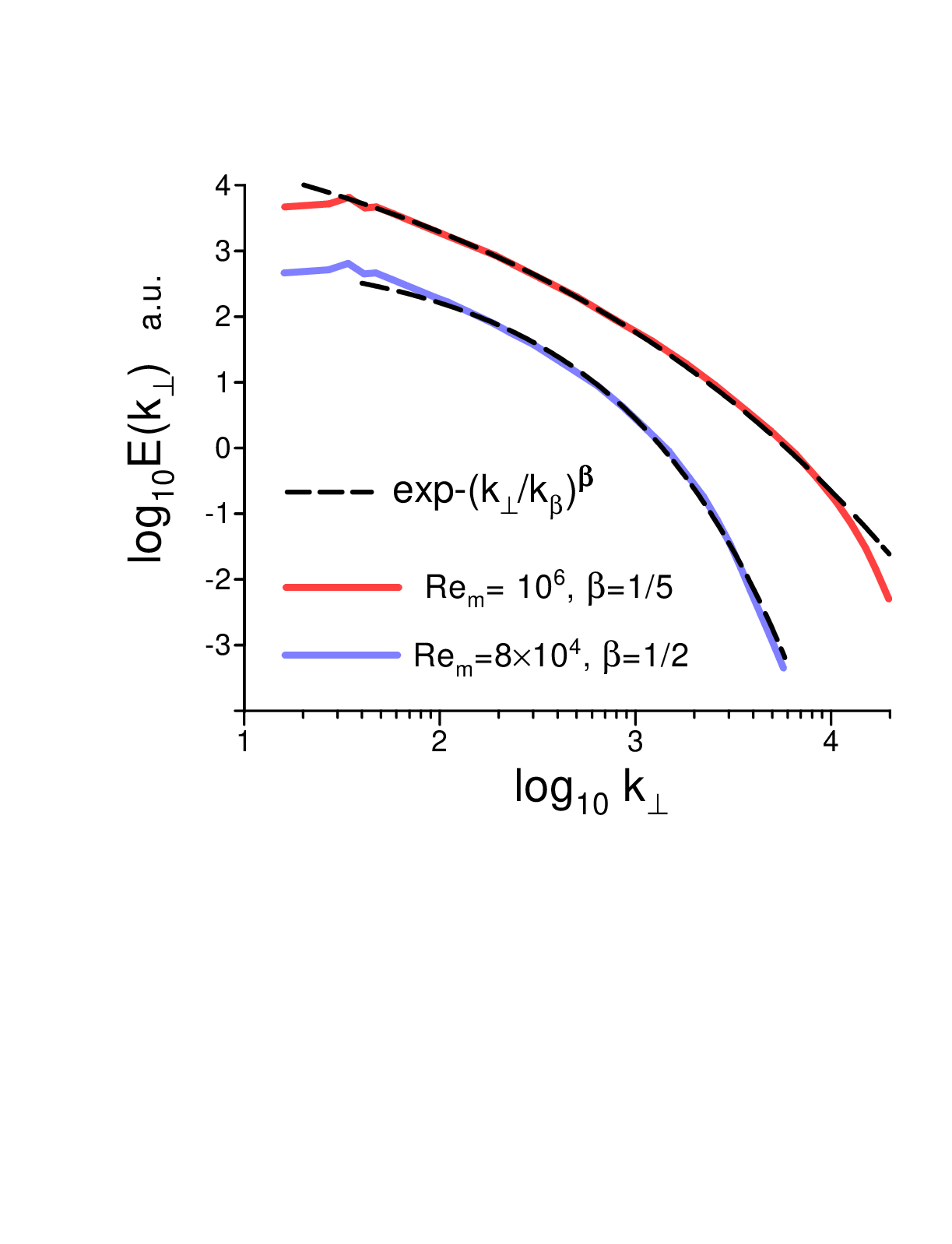} \vspace{-3.75cm}
\caption{DNS computed field perpendicular magnetic energy spectra for two high values of magnetic Reynolds number: $Re_m =8 \times 10^4$ and $Re_m = 10^6$. The spectra are vertically shifted for clarity.} 
\end{figure}
\begin{figure} \vspace{-1.2cm}\centering
\epsfig{width=.45\textwidth,file=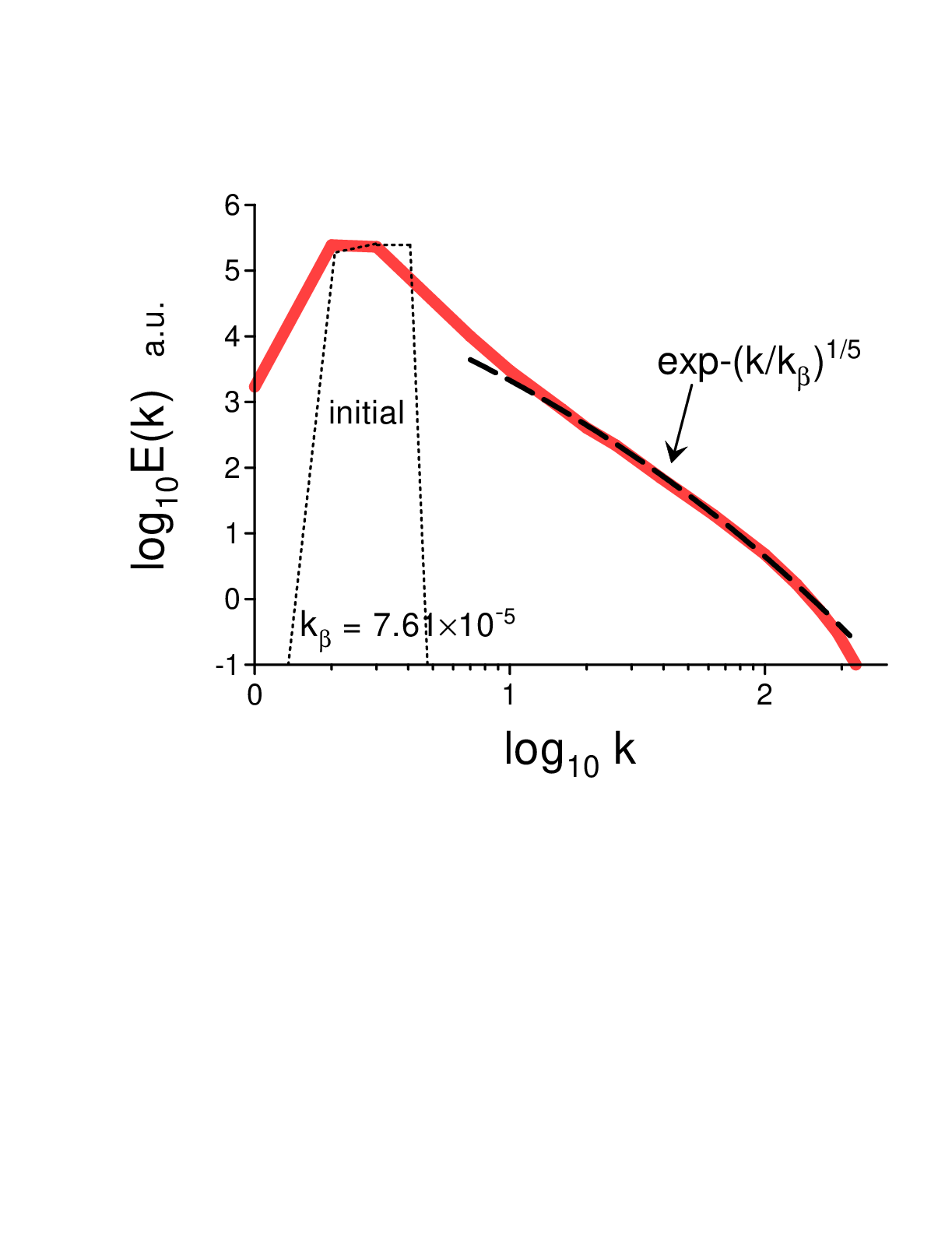} \vspace{-4cm}
\caption{Magnetic energy spectrum of the freely decaying positive EMHD wave packets at $t =0.6$. }
\end{figure}
\begin{figure} \vspace{-1.2cm}\centering
\epsfig{width=.45\textwidth,file=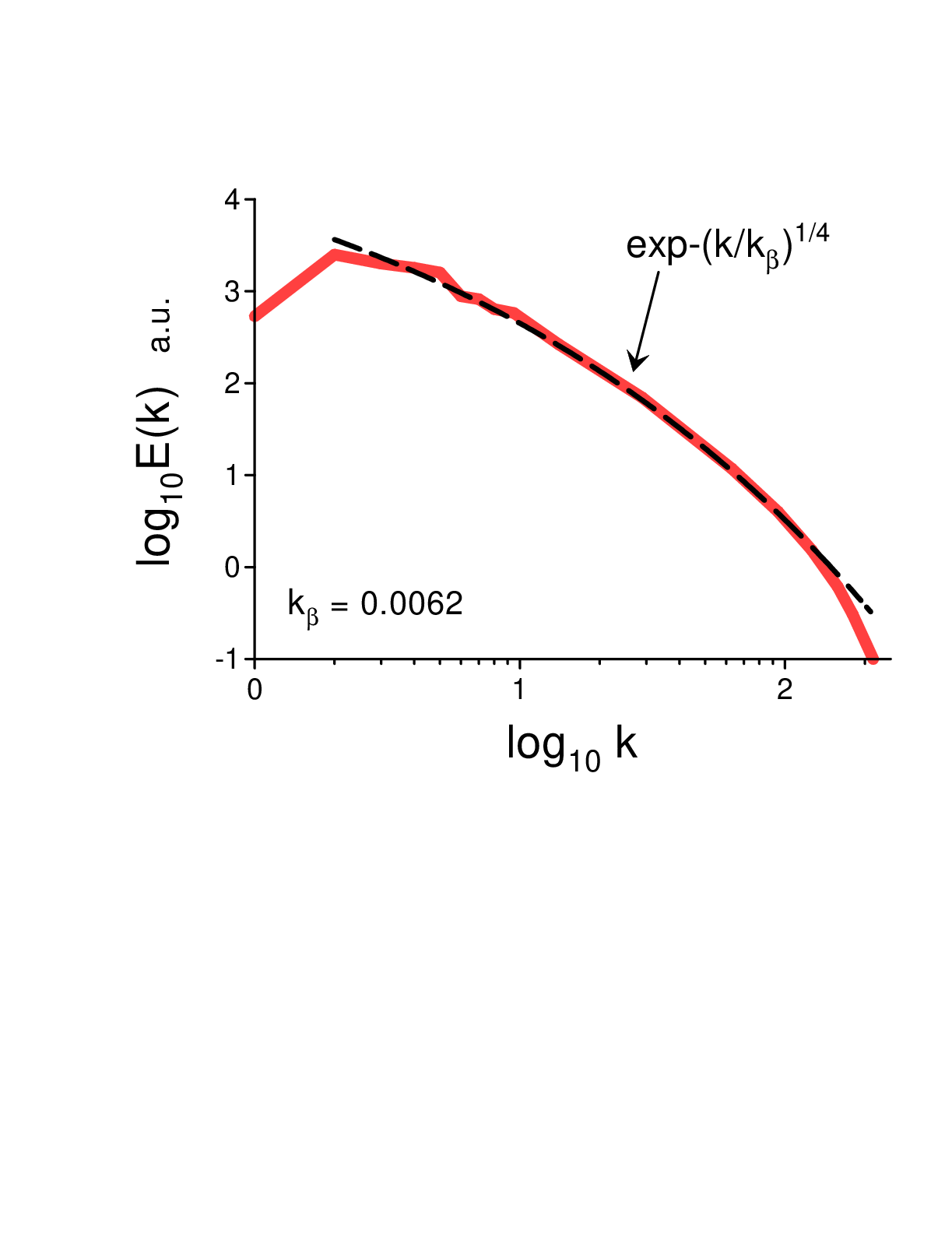} \vspace{-4cm}
\caption{As in Fig. 13 but for negative EMHD wave packets. }
\end{figure}
   In a paper Ref. \cite{mpm} results of direct numerical simulation of a freely decaying MHD flow (with ${\bf f} = 0$ and without the Hall terms in the Eqs. (6-8)) with a superposition of random and deterministic (three ABC flows) initial conditions and considerable mean magnetic helicity were reported. The Kelvin-Helmholtz instabilities with rolling up of current and vorticity sheets were observed at a high Reynolds number ($Re = 10100$). 
   
   Figure 10 shows magnetic energy spectra for $t= 1.6$ in terms of the DNS (the spectral data were taken from Fig. 2b of the Ref. \cite{mpm}). The dashed curve indicates the stretched exponential spectrum Eq. (22).\\
 
     In a paper Ref. \cite{yos} a direct numerical simulation using the forced MHD equations was performed in a 3D spatial box (periodic boundary conditions). The initial fields were randomly generated with kinetic and magnetic energies equal to 0.5 (in the terms of the Ref. \cite{yos}). The initial mean cross helicity was taken almost zero, while the initial mean (normalized) magnetic helicity was rather considerable - 0.515 (at the statistically stationary state the normalized mean magnetic helicity was 0.655). The large-scale random forcing ${\bf f}$ was concentrated in the wavenumber range $1 < k < 2.5$. The magnetic Prandtl number $Pr_m = 1$, the Taylor-Reynolds number $Re_{\lambda} = 150$ and the magnetic Reynolds number $Re_{\lambda}^{(m)} = 306$. \\
    
    Figure 11 shows the magnetic energy spectrum computed for the statistically stationary state using a wavelet method (here $\eta$ is the dissipation scale). The spectral data were taken from Fig. 1 of the Ref. \cite{yos}. The dashed curve indicates correspondence to the stretched exponential spectrum Eq. (22) .\\
     
     In a recent paper Ref. \cite{dong} the MHD equations  
\bea 
\!\!\!\!\!\!\!\partial_t \rho   &=&  - \nabla \cdot (\rho {\bf{u}})  \\
\!\!\!\!\!\!\!\partial_t (\rho {\bf{u}}) + \nabla  \cdot (\rho {\bf{uu}}) \!\!\!\!&=&\!\!\!\!  - \nabla P + \nabla  \cdot ({\bf{BB}}) + \nu {\nabla ^2}(\rho {\bf{u}})  \\ 
\partial_t p + \nabla  \cdot (p {\bf{u}}) &=& (\gamma - 1) \left( { - p \nabla \cdot {\bf{u}} + \eta {\bf{J}}^2} \right) \\
\partial_t {\bf{B}} &=& \nabla  \times ({\bf{u}} \times {\bf{B}} - \eta {\bf{J}})  
\eea

(where  $\gamma$ denotes the adiabatic index) were numerically solved for a very high magnetic Reynolds number $Re_m = 10^6$. The equations Eqs. (24-27) were solved in a periodic spatial box elongated ($1 \times 1 \times 2$) along the direction of an imposed strong mean magnetic field ${\bf B_0}$. The dynamics was initialized by uncorrelated, equipartitioned magnetic and velocity fields. \\

   Figure 12 shows the field perpendicular magnetic energy spectra computed in this DNS for two values of $Re_m = 10^6$ and $Re_m =8 \times 10^4$. The spectral data were taken from Fig. 2 of the Ref. \cite{dong}. The dashed curves indicate correspondence to the stretched exponential spectra: Eq. (12) for $Re_m = 8 \times 10^4$, and Eq. (23) for $Re_m = 10^6$. \\ 
   
   It should be noted that for $Re_m > 10^5$ the authors of the DNS observed a tearing instability of the elongated current sheets resulting in the formation of the small magnetic flux ropes (plasmoids). For $Re_m = 8 \times 10^4$ the flux ropes were practically absent.

\subsection{Electron magnetohydrodynamics (EMHD)}
  
  There are different theoretical (numerical) models for the transitional range between large-scale magnetohydrodynamics and small-scale kinetics (see, for instance, Refs. \cite{sch1},\cite{sch} and references therein). If for the MHD scales one can treat the plasma as a single conducting fluid, for the transitional region a simple two-fluid model - electron magnetohydrodynamics (EMHD), can be applied as a first approximation. This model is rather appropriate for the cases dominated by the magnetic helicity \cite{cho}. It is assumed in the EMHD that ions provide only a motionless charge background for the fast electron flows which carry current  and determine the dynamics of the magnetic field ${\bf B}$
\bea
 {\bf u}_e  = -\frac{{\bf J}}{n_e e} &=& -\frac{c}{4\pi n_e e} [\nabla \times {\bf B}]   
\eea
where ${\bf J}$ is the electric current density, ${\bf u_e}$ is the electron velocity,  $n_e$ is the electron number density \cite{kin}.

   Substitution of the Eq. (28) into the equation
\bea
\frac{\partial {\bf B}}{\partial t} = \nabla \times [{\bf u}_e \times {\bf B}] + \eta \nabla^2 {\bf B}       
\eea
provides an equation describing the dynamics of the magnetic field in the incompressible EMHD 
\bea
\frac{\partial {\bf B}}{\partial t} = -\frac{c}{4\pi n_e e} \nabla \times [(\nabla \times {\bf B}) \times {\bf B}] +   \eta \nabla^2 {\bf B}   
\eea

\begin{figure} \vspace{-0.08cm}\centering
\epsfig{width=.5\textwidth,file=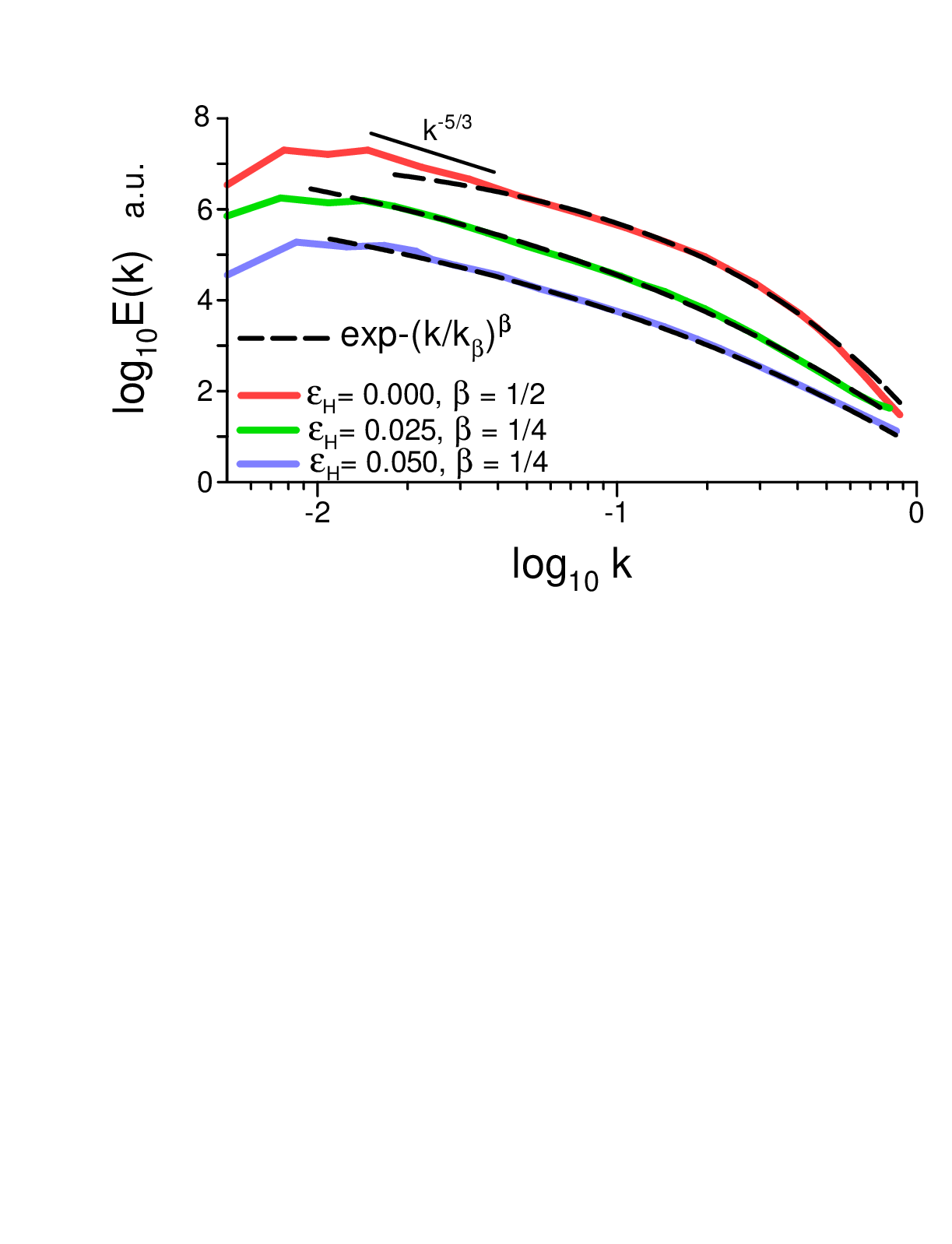} \vspace{-6.55cm}
\caption{Magnetic magnetic energy spectra for the decaying isotropic MHD and Hall MHD for different values of the Hall parameter. The spectra are vertically shifted for clarity.}
\end{figure}

  It should be noted that magnetic helicity for the EMHD is conserved in the ideal case similar to the MHD \cite{sch1}. Unlike the MHD Alfv{\`e}n waves the EMHD waves are helical and carry a nonzero magnetic helicity. Also, the EMHD waves are dispersive and can interact with each other. In a strong imposed magnetic field the direction of propagation of the EMHD wave packets is either parallel or antiparallel to the magnetic field lines. One can speak about `positive' and `negative' wave packets in respect to the sign of the magnetic helicity carried by the wave packets. \\

  In the Ref. \cite{cho} the Eq. (30) was solved numerically in a periodic box for the decaying case with an imposed uniform magnetic filed ${\bf B}_0$ (so that ${\bf B} = {\bf B}_0 + {\bf b}$).  Figure 13 shows the power spectrum of magnetic energy of the freely decaying positive EMHD wave packets obtained with this DNS (the spectral data were taken from Fig. 4 of the Ref. \cite{cho} for $t = 0.6$ in terms of the Ref. \cite{cho}). The dotted lines show the initial energy spectrum (only positive waves were present in the DNS at $t =0$). The dashed curve indicates the stretched exponential spectrum Eq. (23).\\  
  
   The EMHD wave packets moving in one direction interact with each other and, therefore, can generate counterpropagating waves. Figure 14 shows the power spectrum of magnetic energy of the freely decaying {\it negative} EMHD wave packets obtained with this DNS (the spectral data were taken from Fig. 4 of the Ref. \cite{cho} for $t = 0.6$). The dashed curve indicates the stretched exponential spectrum Eq. (22).\\  

\subsection{Hall magnehydrodynamics} 

   In a paper Ref. \cite{ma} results of direct numerical simulations of the decaying, incompressible, homogeneous, isotropic MHD and Hall MHD were reported. The DNS was performed in a periodic spatial cube with random phase initial conditions and initial energy spectrum
\bea
E(k) \propto k^4 \exp-(k^2/k_0^2)    
\eea
for both the velocity and magnetic field (here $k_0 =2$). The initial kinetic and magnetic energies were approximately the same.\\

  Figure 15 shows the DNS computed magnetic energy spectra for the decaying isotropic MHD and Hall MHD for different values of the Hall parameter $\epsilon_H =d_i/L_0$ at $t =2.4$ in the DNS terms. The spectral data were taken from Fig. 2b of the Ref. \cite{ma}.  The dashed curves indicate the stretched exponential spectra: Eq. (12) for $\epsilon_H = 0$ (MHD), and Eq. (22) for  $\epsilon_H = 0.025$, and $\epsilon_H = 0.050$ for Hall MHD.\\

\begin{figure} \vspace{-0.5cm}\centering
\epsfig{width=.5\textwidth,file=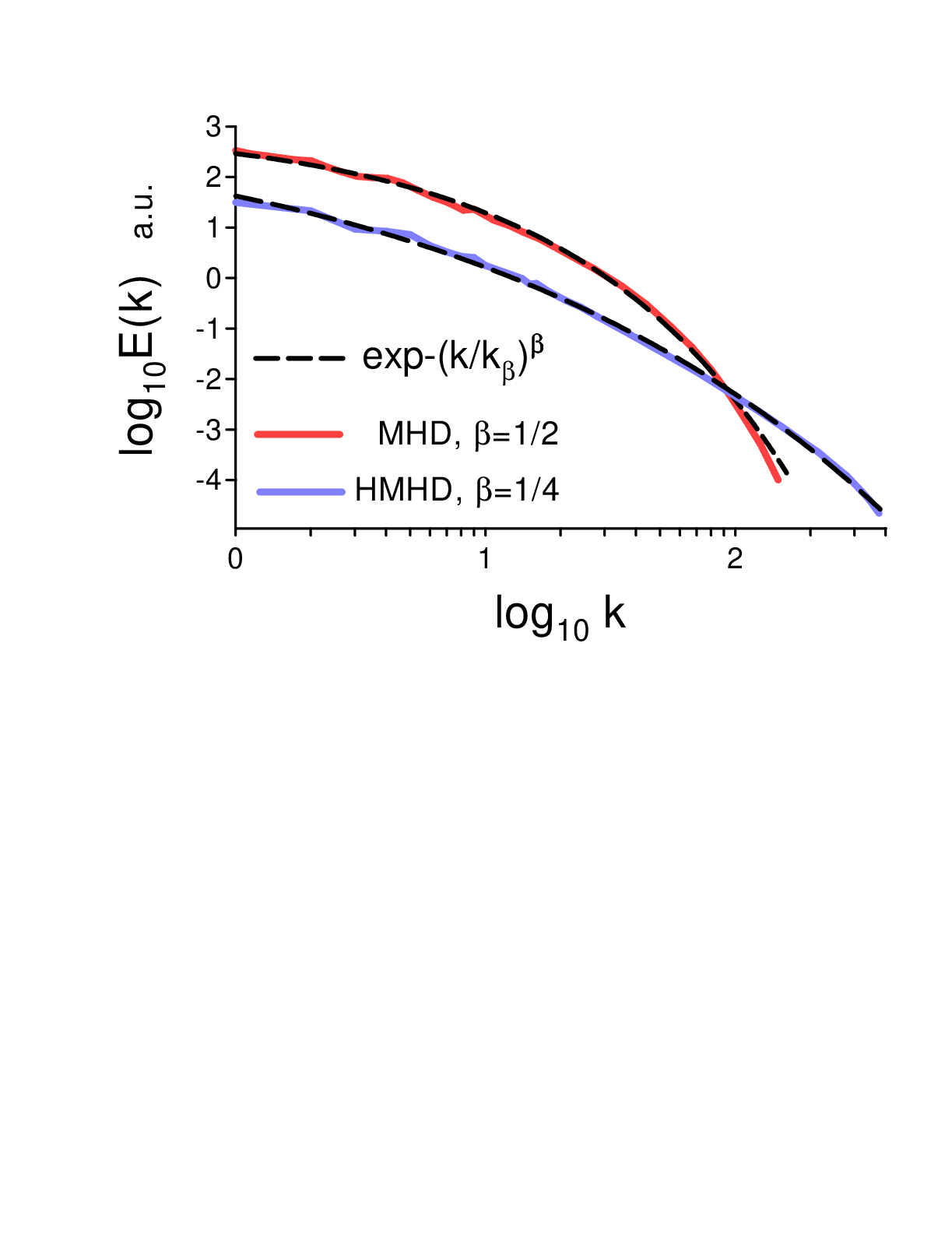} \vspace{-6.13cm}
\caption{Magnetic magnetic energy spectra for the weakly incompressible decaying isotropic MHD and Hall MHD. The spectra are vertically shifted for clarity.}
\end{figure}
 
  In a recent paper Ref. \cite{fold} results of a Lattice Boltzmann numerical simulation of a weakly compressible MHD and Hall MHD were reported.  The simulations were performed in a periodic cubic lattice and were initialized with the Orszag-Tang vortex. \\
  
   Figure 16 shows the magnetic energy spectra computed at the peak of magnetic dissipation both for the MHD case and for the Hall MHD case ($\epsilon_H =0.015$, $Re =6000$, $Pr_m =1$). The spectral data were taken from Fig. 11 of the Ref. \cite{fold} for the MHD case and from Fig. 14 of the Ref. \cite{fold} for the Hall MHD case. 
   
    The dashed curves indicate the stretched exponential spectra: Eq. (12) for the MHD case, and Eq. (22) for the Hall MHD case.
   
\subsection{A fully kinetic numerical simulation}

\begin{figure} \vspace{-0.5cm}\centering
\epsfig{width=.5\textwidth,file=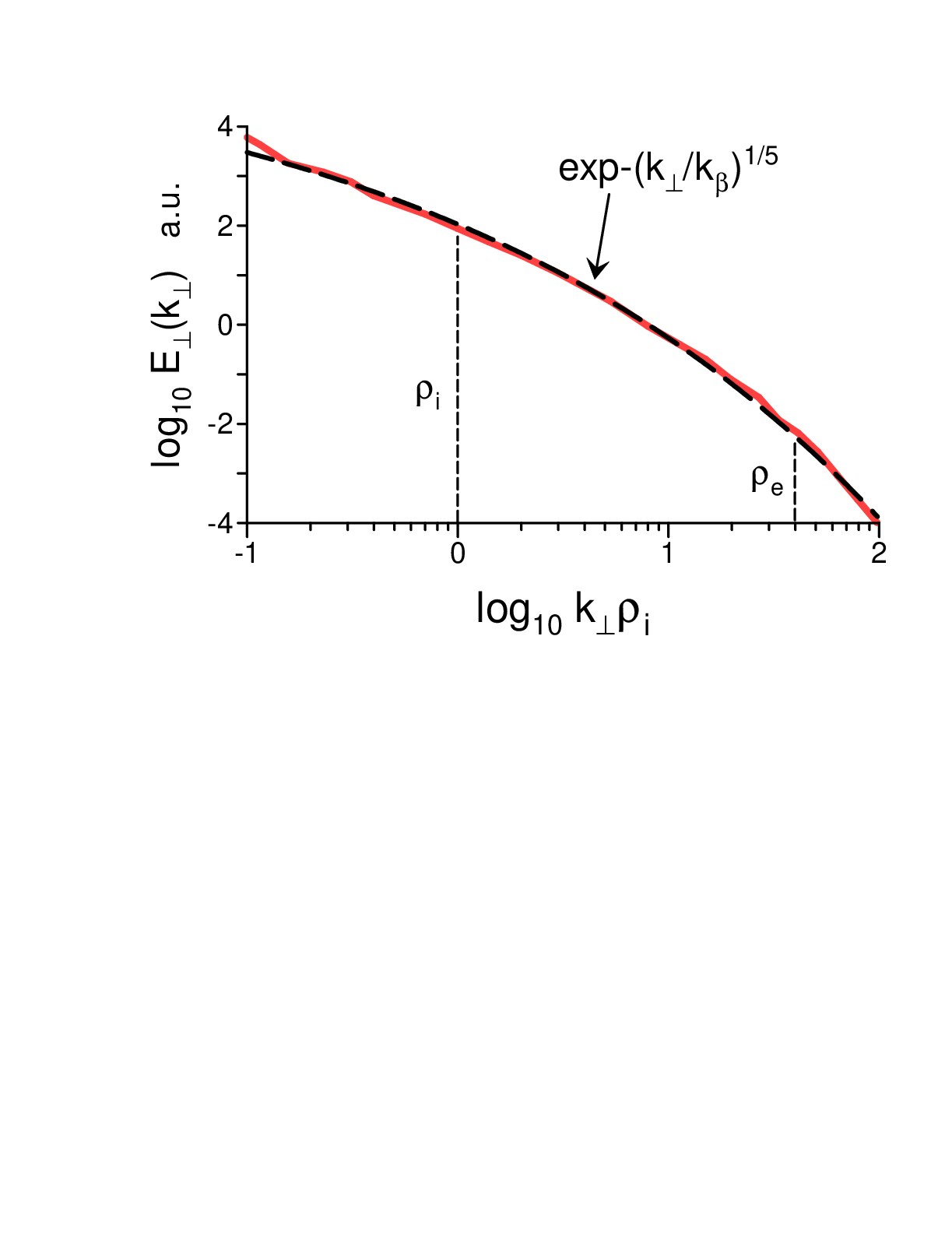} \vspace{-6.1cm}
\caption{ Perpendicular magnetic energy spectrum vs $k_{\perp} \rho_i$ ($\rho_i$ and $\rho_e$ are ion and electron gyroradius respectively) for a fully kinetic numerical simulation.}
\end{figure}

\begin{figure} \vspace{-0.5cm}\centering
\epsfig{width=.45\textwidth,file=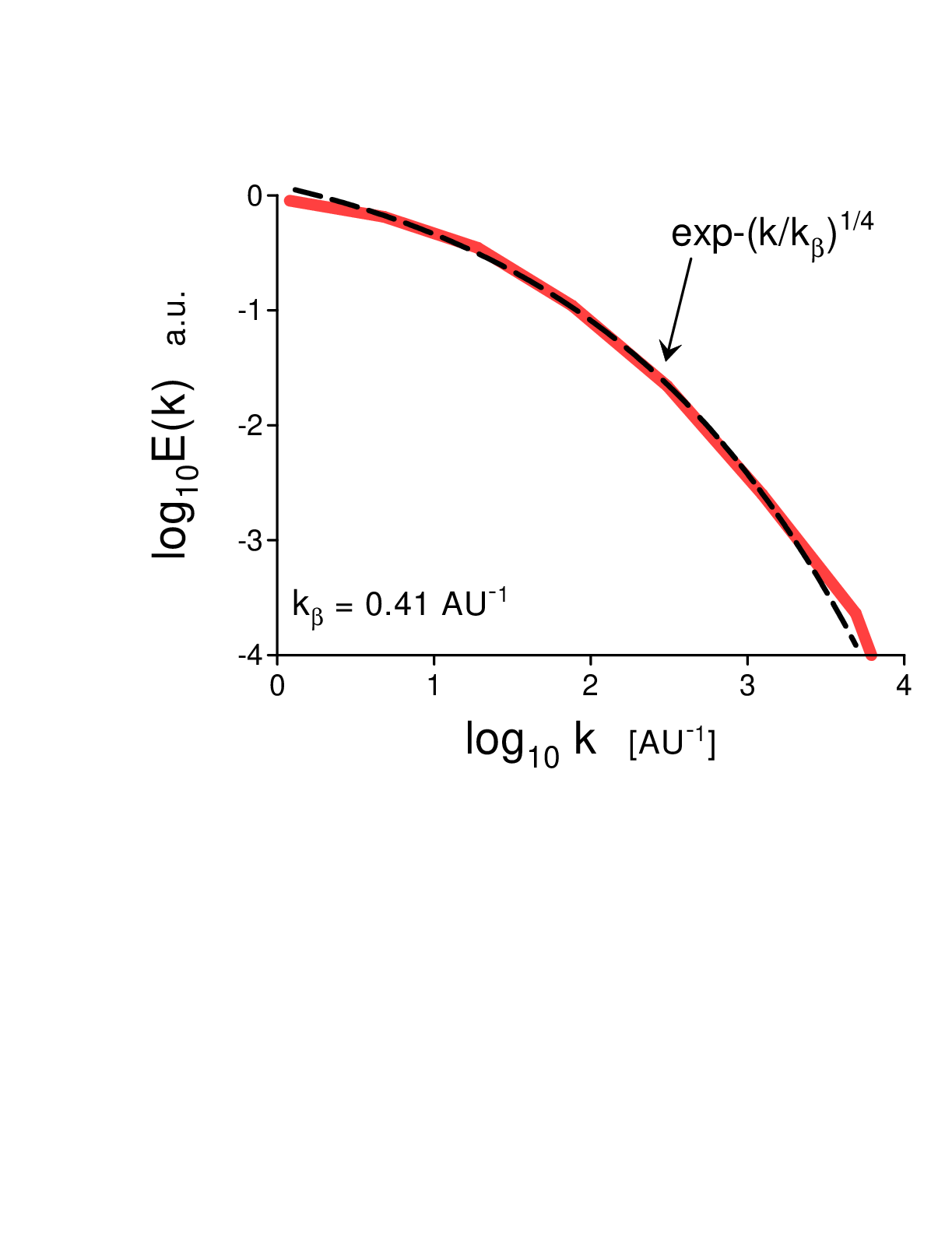} \vspace{-4.03cm}
\caption{Total magnetic energy spectrum for the fast solar wind at $2.8 < R < 4.5$AU (Ulysses, 1993-1996yy).}
\end{figure}

\begin{figure} \vspace{-1.4cm}\centering
\epsfig{width=.45\textwidth,file=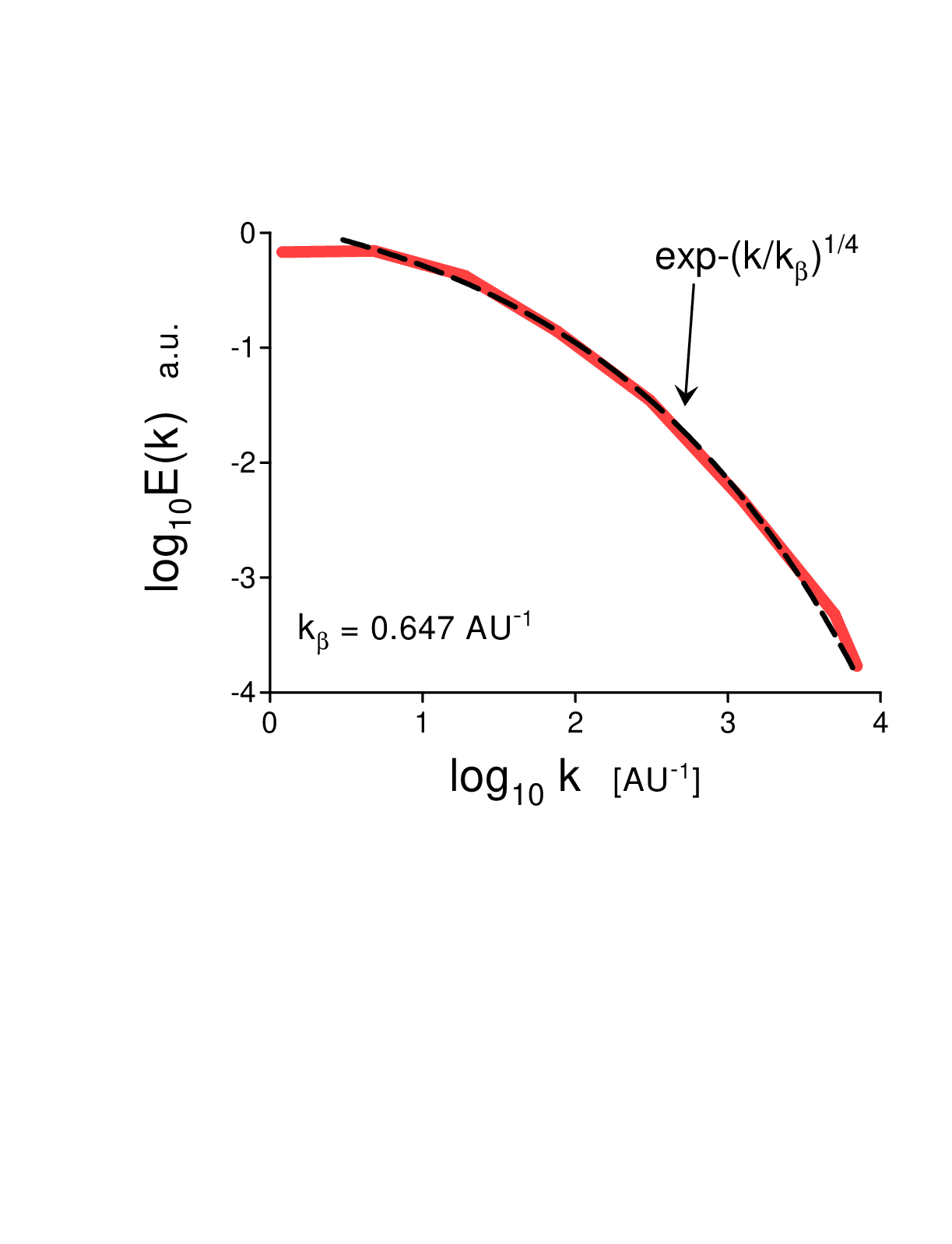} \vspace{-3.65cm}
\caption{The same as in Fig. 18 but at $1.5 < R < 2.8$ AU.}
\end{figure}

 In a paper Ref. \cite{thd} results of a gyrokinetic numerical simulation were reported. In this simulation the authors numerically solved the electromagnetic gyroaveraged Vlasov–Maxwell equations for electrons and protons as well as for the component of the vector potential ${\bf A}$ parallel to the equilibrium magnetic field ${\bf B}_0$, for the parallel perturbation of the magnetic field, and the electrostatic potential. The equilibrium distribution functions for protons and electrons are Maxwellian. A conservative collision operator was linearized and gyroaveraged. The simulation was driven by an oscillating Langevin antenna which was coupled to the component of the vector potential parallel to ${\bf B}_0$. 
 The parameters of the plasma were chosen as follows: $\beta_i = 1$, $m_i/m_e = 1836$, and $Ti/Te = 1$. The simulation was performed in a periodic three-dimensional spatial domain elongated along the uniform magnetic field ${\bf B}_0$.\\
 
   Figure 17 shows the perpendicular magnetic energy spectrum vs $k_{\perp} \rho_i$ ($\rho_i$ and $\rho_e$ are ion and electron gyroradius respectively). The spectral data were taken from Fig. 1 of the Ref. \cite{thd}. The dashed curve indicates the stretched exponential spectrum Eq. (23). The short-dashed vertical lines are drawn to indicate the positions of the gyroradii $\rho_i$ and $\rho_e$.

\section{Solar wind}

  Solar wind is characterized by a wide variety of physical conditions. Distance from the Sun, fast or slow wind, etc. determine these physical conditions. Spacecraft measurements provide us with good-quality data observed in this unique natural laboratory. \\
  
   Figure 18, for instance, shows the total magnetic energy spectrum calculated using measurements made by Ulysses mission for the 1993-1996yy period at the high heliolatitudes and in the fast solar wind. The spectral data were taken from Fig. 3 of the Ref. \cite{bran} for $2.8 < R < 4.5$ astronomical units (AU), where $R$ is the distance from the Sun. The spectral data were rescaled by a factor $4\pi R^2$ before averaging. Figure 19 shows the analogous spectrum for $1.5 < R <2.8$ AU (the spectral data were taken from the Fig. 3 of the Ref. \cite{bran}). The wavenumber spectra were obtained using Taylor's ``frozen in'' hypothesis. The dashed curves indicate correspondence to the stretched exponential spectrum Eq. (22).\\
   
   Figure 20 shows the total magnetic energy spectrum for the fast solar wind at $R =1.4$AU computed using measurements made by the Ulysses mission. The spectral data were taken from Fig. 1 of the Ref. \cite{tbt} (see also Refs. \cite{bc},\cite{bt}). The dashed curves indicate correspondence to the stretched exponential spectrum Eq. (23) for the large spatial scales (small frequencies in the figure) and to the stretched exponential spectrum Eq. (22) for the small spatial scales (large frequencies in the figure).\\
   
     Figure 21 shows the power spectrum of magnetic field fluctuations measured at  $R = 1$AU by ARTEMIS-P2 spacecraft. The spectral data were taken from Fig. 1 of Ref. \cite{chen}. 
 The dashed curve indicates correspondence to the stretched exponential spectrum Eq. (23). The ion gyroradius $\rho_i$ and ion inertial length $d_i$ are indicated in Fig. 21 by the vertical dotted lines. The appearance of the Kolmogorov-like power law spectrum at large scales is indicated by the straight solid line with the slope -5/3 (cf. Ref. \cite{bs}). \\
\begin{figure} \vspace{-0cm}\centering \hspace{-1cm}
\epsfig{width=.49\textwidth,file=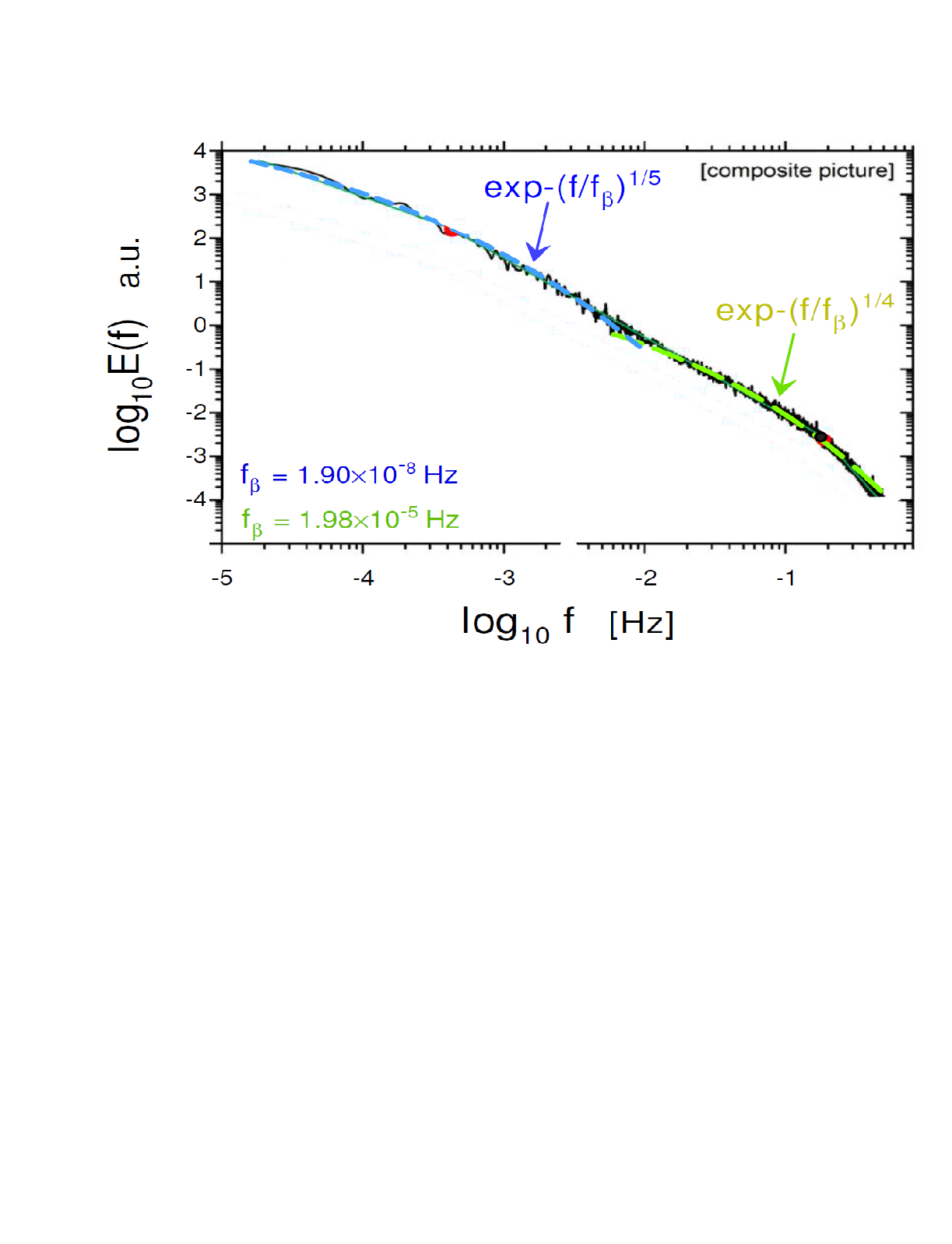} \vspace{-5.55cm}
\caption{Total magnetic energy spectrum for the fast solar wind at $R =1.4$AU (Ulysses).}
\end{figure}
\begin{figure} \vspace{-0.5cm}\centering
\epsfig{width=.47\textwidth,file=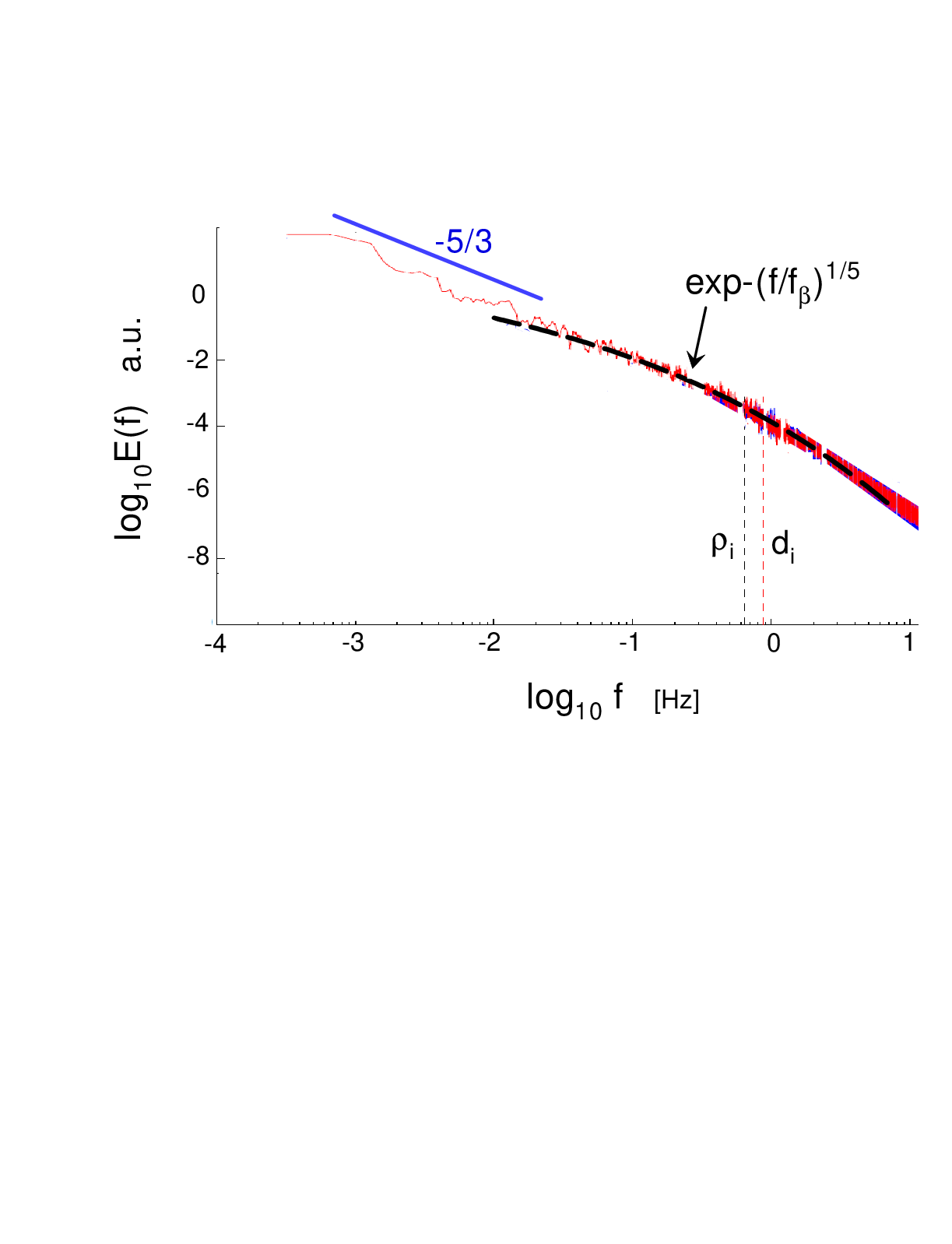} \vspace{-4.6cm}
\caption{Power spectrum of magnetic field fluctuations for the fast solar wind at $R = 1$AU (ARTEMIS-P2). }
\end{figure}
\begin{figure} \vspace{-0.4cm}\centering
\epsfig{width=.46\textwidth,file=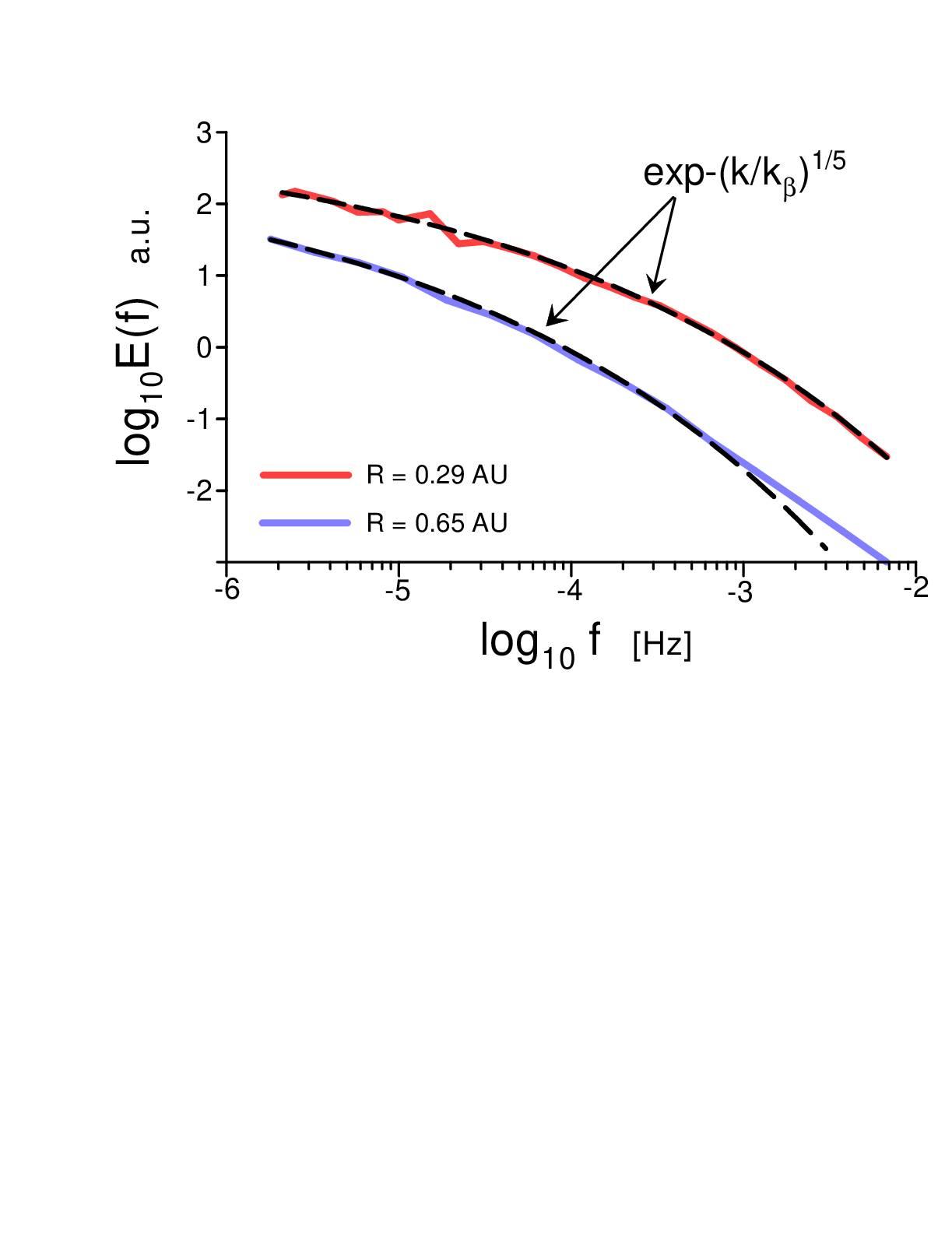} \vspace{-5.05cm}
\caption{Total magnetic energy spectrum for the fast solar wind at $R = 0.65$AU and $R = 0.29$AU (Helios 2). }
\end{figure}
\begin{figure} \vspace{-0.1cm}\centering
\epsfig{width=.43\textwidth,file=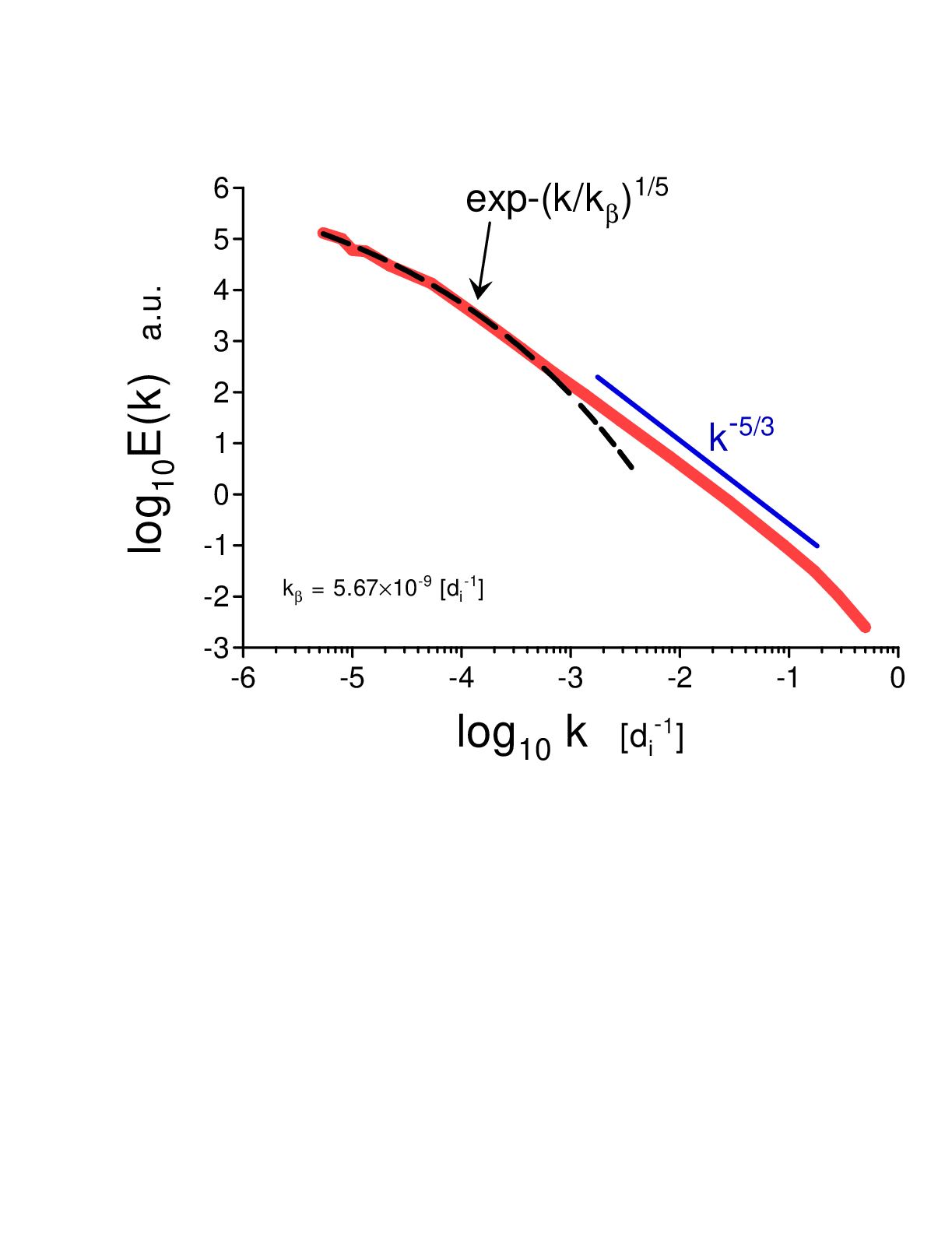} \vspace{-3.95cm}
\caption{Magnetic energy spectrum computed from the auto-correlation of the magnetic field fluctuations measured at $R = 0.16$AU (near Parker Solar Probe first perihelion). }
\end{figure}
 
  Space probe Helios 2 observed solar wind in the ecliptic and much closer to the Sun \cite{rosen}. Figure 22 shows the total magnetic energy spectrum for the fast solar wind at $R= 0.65$AU and $R =0.29$AU. The spectral data were taken from Fig. 1 of Ref. \cite{tbt}  (see also Refs. \cite{bc},\cite{bt}). The dashed curves indicate correspondence to the stretched exponential spectrum Eq. (23).\\

 The NASA Parker Solar Probe studies the deeper inner heliosphere \cite{fox}. Figure 23 shows the magnetic energy spectrum computed from the auto-correlation of the magnetic field fluctuations measured at $R = 0.16$AU (near Parker Solar Probe's first perihelion). The spectral data were taken from Fig. 1b of the Ref. \cite{gue}. Taylor’s hypothesis was used to convert the time lags $\tau$ into the wavenumbers (in this figure $k = 1/V_0\tau$, where $V_0$ is the mean solar wind speed), and the wavenumber was taken in the units of the inverse ion inertial length $1/d_i$. The dashed curve indicates correspondence to the stretched exponential spectrum Eq. (23) for the large spatial scales. For the small spatial scales, a Kolmogorov-like spectrum was observed (indicated by the straight line in the log-log scales).

\section{In the Earth's magnetosphere}
 
\subsection{Magnetosheath}
 
   Let us start with the Earth's magnetosheath, i.e. the area of the magnetosphere bounded by the magnetopause and the bow shock. 
   
  Figure 24 shows the power spectrum of magnetic fluctuations in the magnetosheath near the bow shock, whereas Fig. 25 shows the power spectrum of magnetic fluctuations in the magnetosheath near the magnetopause (toward the flank region). The spectral data were taken from Fig. 1 of a paper Ref. \cite{huang2} (the raw data were obtained by the Cluster mission, see Ref. \cite{huang2} for more details and references). The dashed curve in Fig. 24 indicates the spectral law Eq. (23), whereas the dashed straight line in Fig. 25 indicates the Kolmogorov-like spectral law (the authors of the Ref. \cite{huang2} argued that the nature of this Kolmogorov-like spectral law is still unclear). The ion gyroradius $\rho_i$ is indicated by the vertical dotted lines.\\      
  
\begin{figure} \vspace{-0cm}\centering
\epsfig{width=.45\textwidth,file=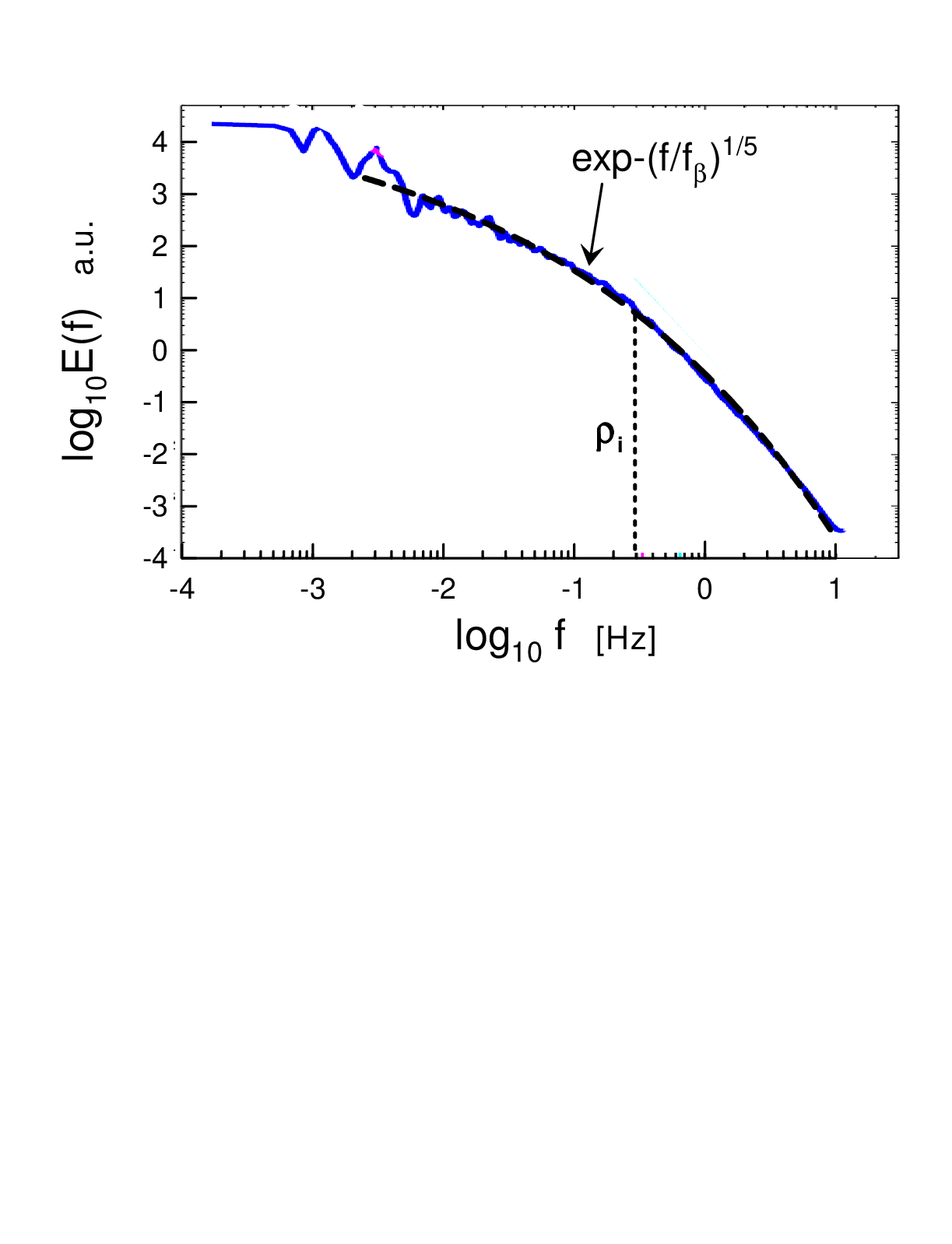} \vspace{-4.95cm}
\caption{Power spectrum of magnetic fluctuations in the magnetosheath near the bow shock. Cluster spacecraft measurements.}
\end{figure}
\begin{figure} \vspace{+0cm}\centering
\epsfig{width=.47\textwidth,file=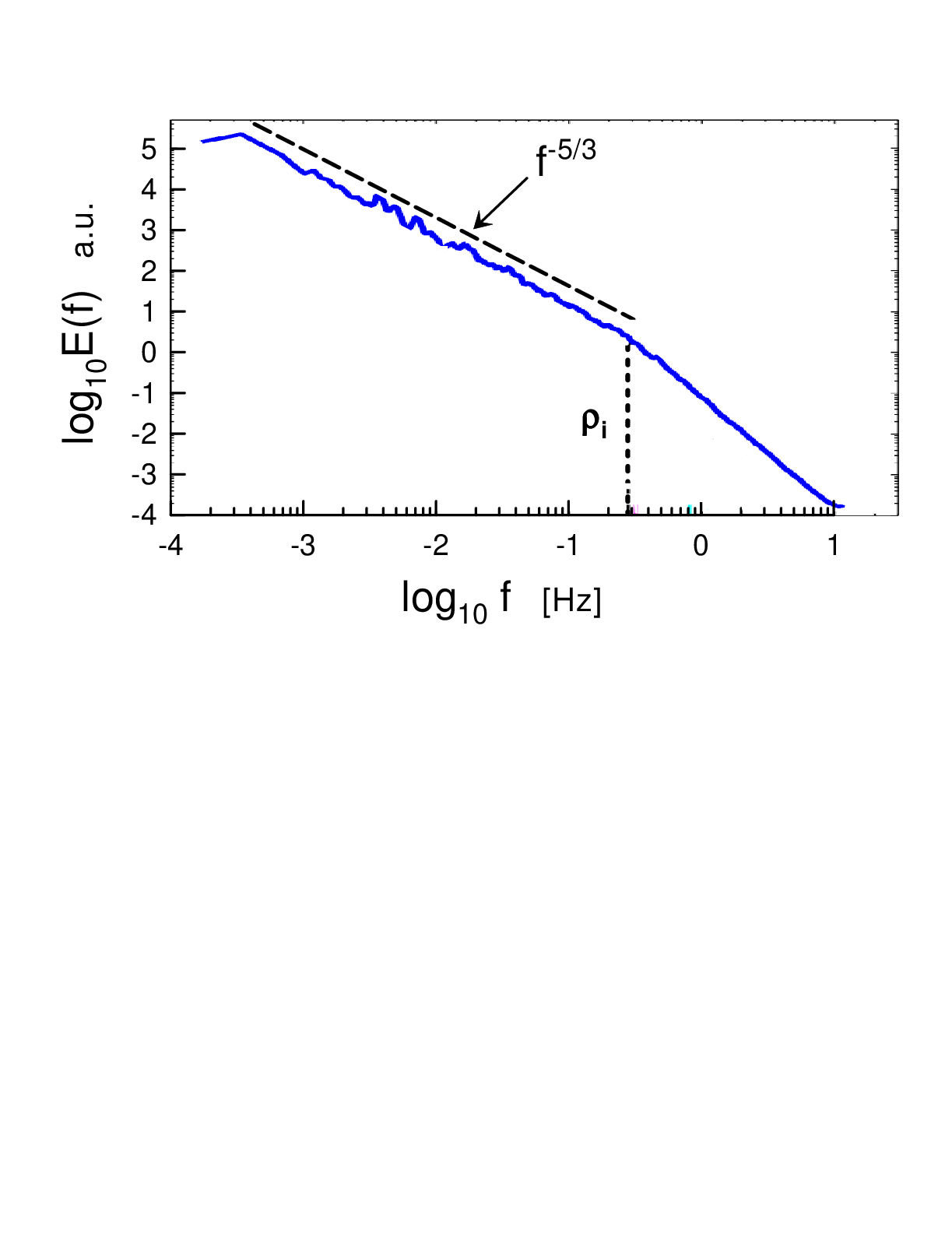} \vspace{-5.45cm}
\caption{Power spectrum of magnetic fluctuations in the magnetosheath near the magnetopause (toward the flank region). Cluster spacecraft measurements.}
\end{figure}
\begin{figure} \vspace{-0.47cm}\centering
\epsfig{width=.45\textwidth,file=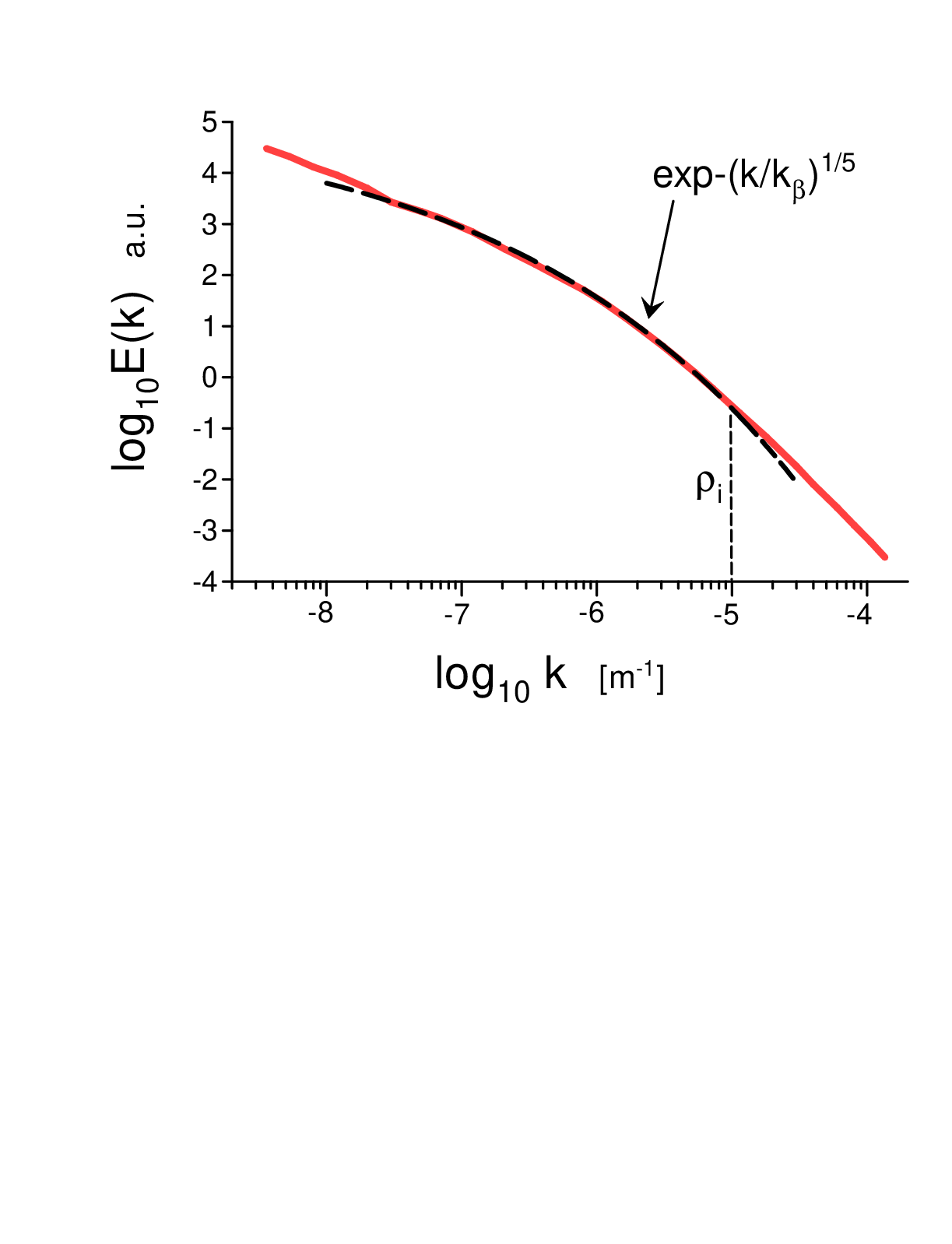} \vspace{-4.7cm}
\caption{Power spectrum of magnetic fluctuations in the magnetosheath at 14.8 $R_E$ (Magnetospheric Multi-Scale spacecraft - MMS).}
\end{figure}
\begin{figure} \vspace{-0.6cm}\centering
\epsfig{width=.45\textwidth,file=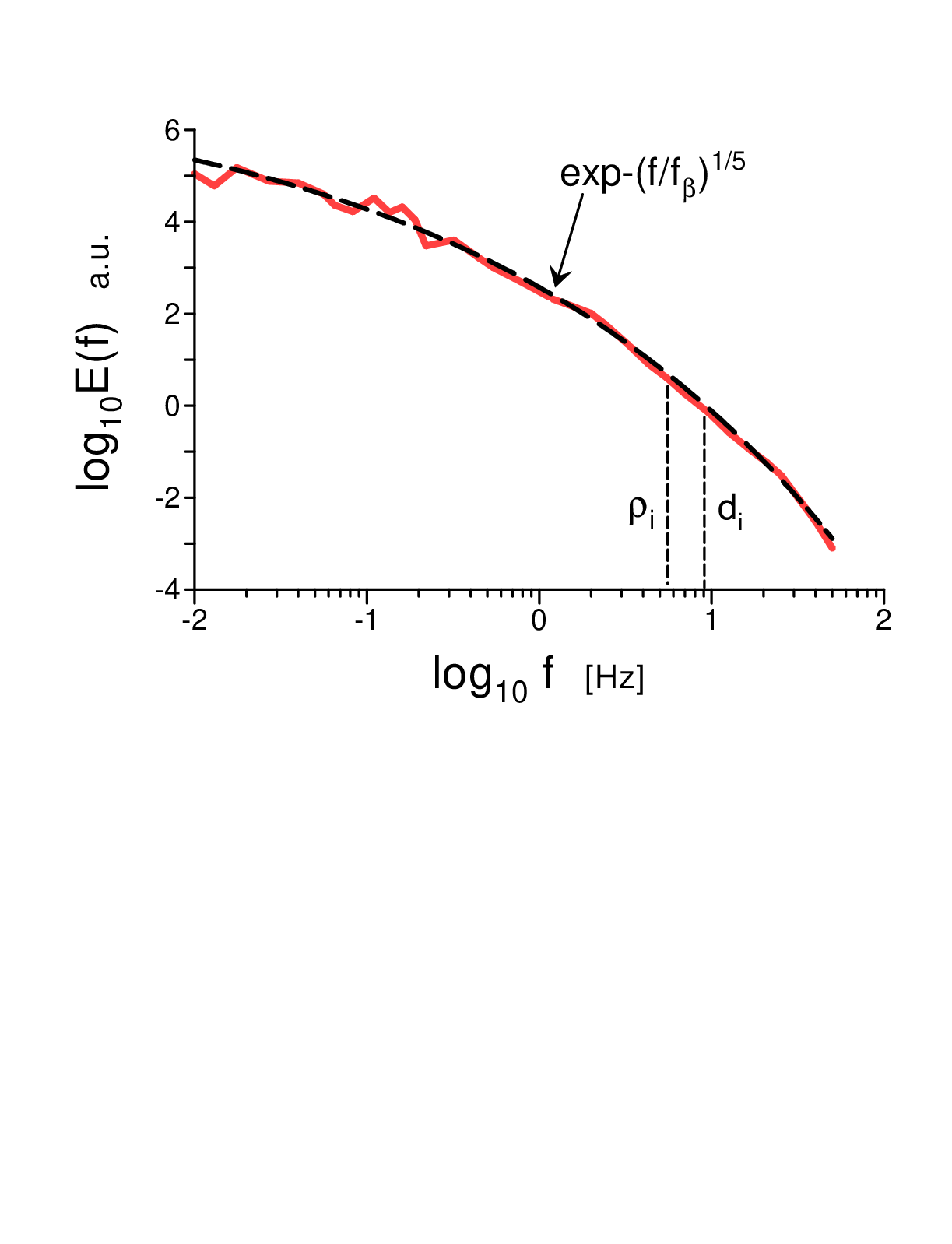} \vspace{-4.6cm}
\caption{Power spectrum of magnetic fluctuations in the magnetosheath (MMS). The magnetosheath event under consideration is believed to be reconnecting intense current sheets.}
\end{figure}

  Figure 26 shows the power spectrum of magnetic fluctuations in the magnetosheath computed for the data obtained by the Magnetospheric Multi-Scale spacecraft (the spectral data were taken from Fig. 1 of the Ref. \cite{par}). Taylor's hypothesis was used to transform the frequency spectrum into the wavenumber one (the authors of the Ref. \cite{par} provide certain arguments supporting the application of the Taylor hypothesis to this case). The spacecraft was approximately at 14.8 $R_E$ ($R_E$ is the Earth’s radius). The dashed curve in Fig. 26 indicates the spectral law Eq. (23).\\ 
  
  Figure 27 shows the power spectrum of magnetic fluctuations in the magnetosheath computed for the data obtained by the Magnetospheric Multi-Scale spacecraft (the spectral data were taken from Fig. 2 of a recent paper Ref. \cite{man}). The magnetosheath event under consideration is believed to be reconnecting intense current sheets. The dashed curve in Fig. 27 indicates the spectral law Eq. (23).\\

 \subsection{Reconnection jet in the plasma sheet of the Earth's magnetosphere}
 
\begin{figure} \vspace{-0.7cm}\centering \hspace{-1cm}
\epsfig{width=.5\textwidth,file=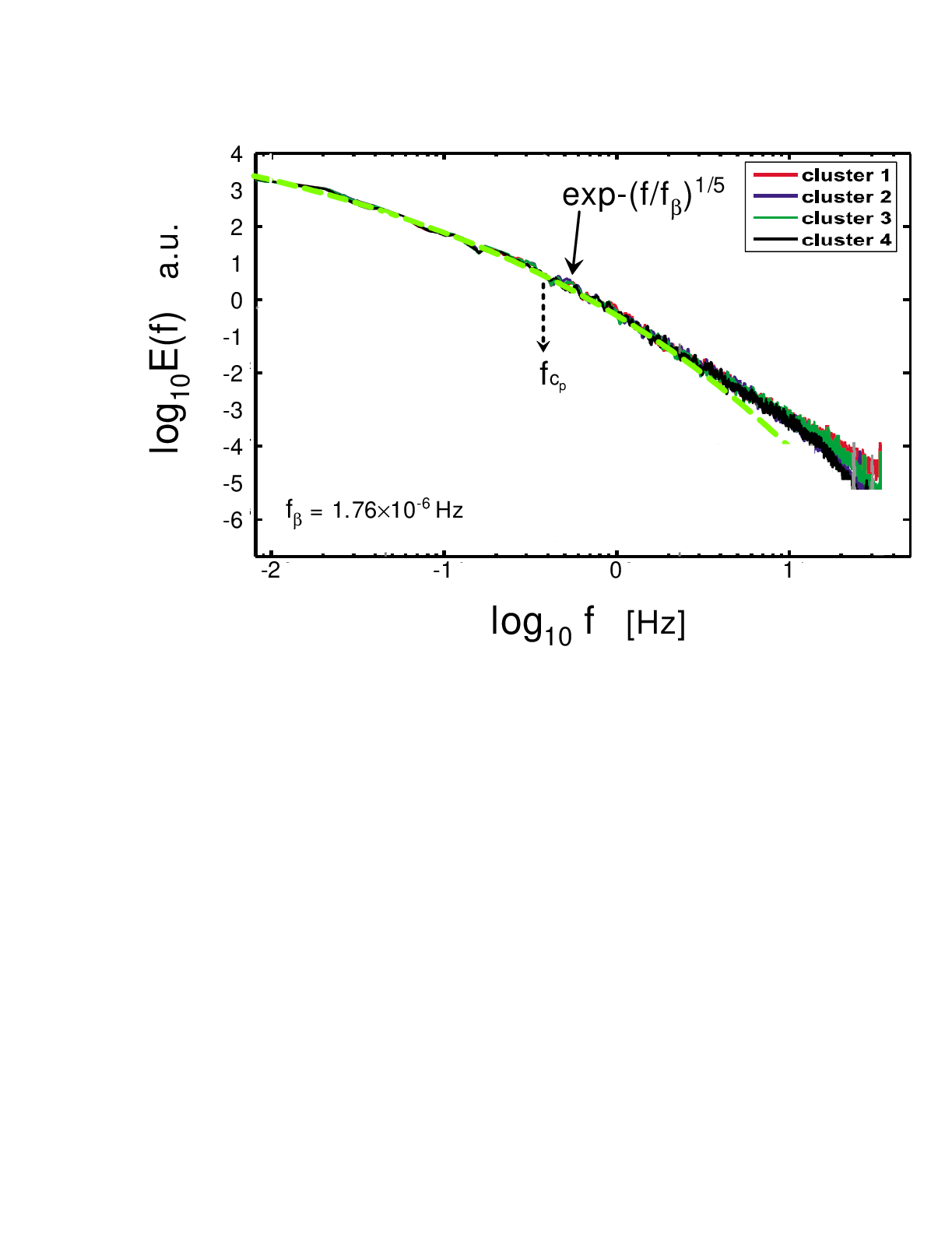} \vspace{-5.55cm}
\caption{Power spectrum of the magnetic field measured in the magnetospheric reconnection jet (Cluster mission). }
\end{figure}
  
\begin{figure} \vspace{-1cm}\centering
\epsfig{width=.45\textwidth,file=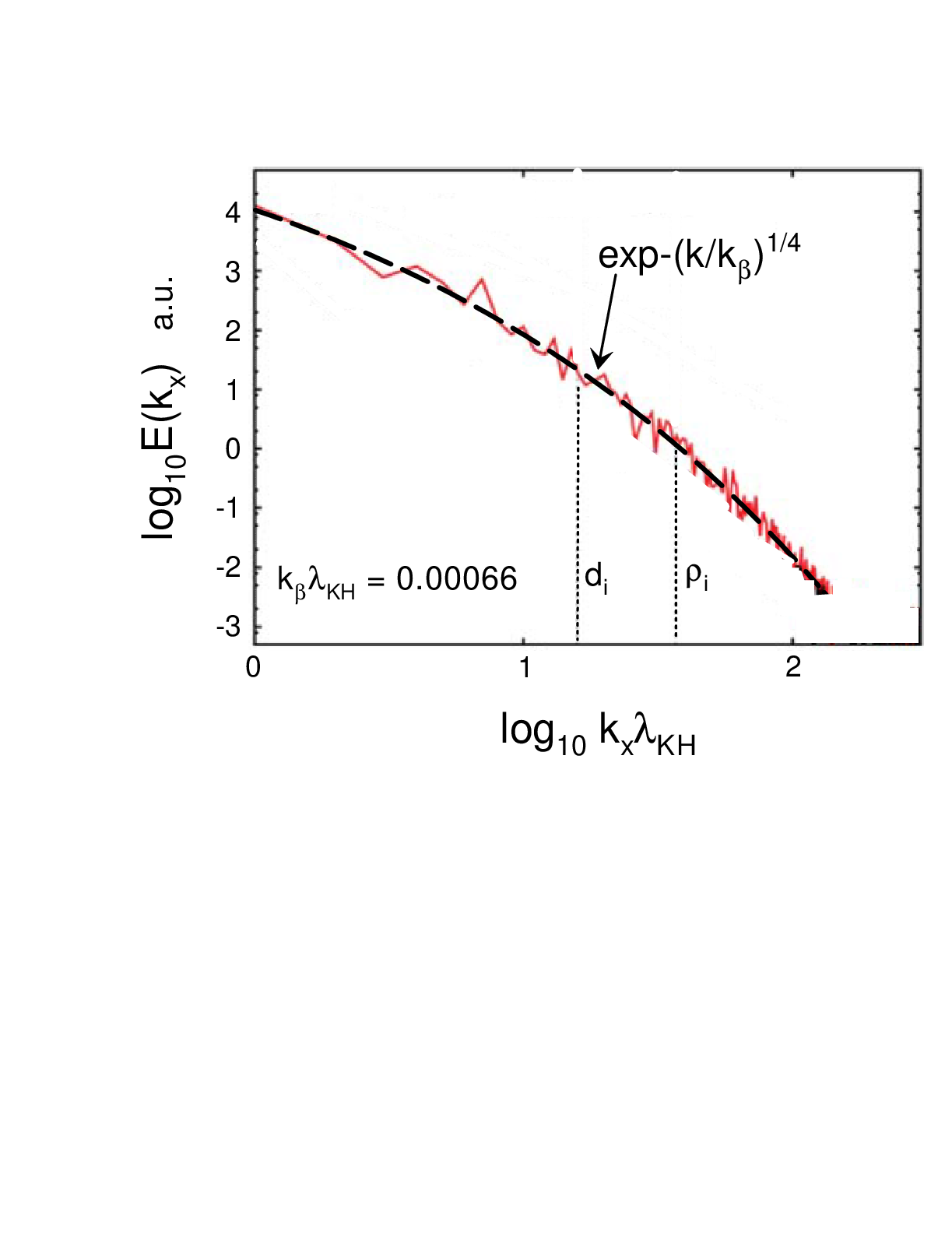} \vspace{-4.1cm}
\caption{Power spectrum of a magnetic field component computed in the numerical simulation and corresponding to the magnetic field component normal to the nominal magnetopause in the MMS observations.}
\end{figure}
\begin{figure} \vspace{-0.5cm}\centering
\epsfig{width=.45\textwidth,file=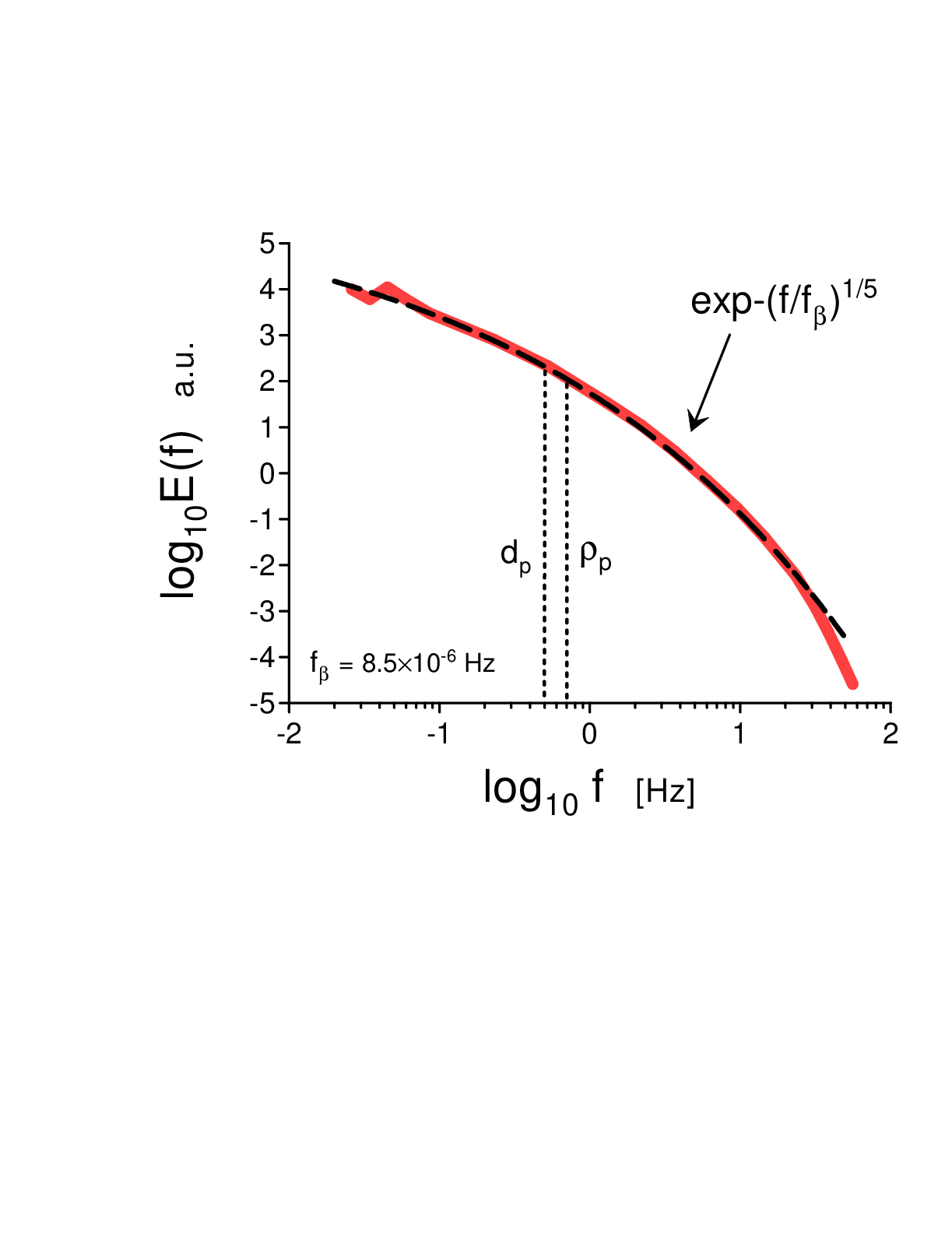} \vspace{-3.6cm}
\caption{Power spectrum of the magnetic field component normal to the nominal magnetopause measured by the MMS for the Kelvin-Helmholtz active period. }
\end{figure}

  In a paper Ref. \cite{huang} results of measurements with four Cluster spacecraft in the plasma sheet of the Earth's magnetosphere were reported. The event under consideration was a high-speed reconnection jet with a moderate mean (guide) magnetic field (see Introduction). Based on the observations the authors of the Ref. \cite{huang} argued that the observed turbulence was strongly driven by the reconnection jet (cf Ref. \cite{alaoui}).
  
  Figure 28 shows the power spectrum of the magnetic field measured during the event. The spectral data were taken from Fig. 1 of the Ref. \cite{huang}. The position of the proton cyclotron frequency $f{\tiny c}_p$ is indicated by the vertical dotted arrow. The dashed curve indicates the stretched exponential spectrum Eq. (23) (cf Fig. 9 and corresponding comments).\\
 
 \subsection{Kelvin-Helmholtz vortices at the Earth’s magnetopause} 
 
 The Kelvin-Helmholtz vortices (or the flow shear-driven Kelvin-Helmholtz instability) at Earth’s magnetopause can be a significant factor for the transport of momentum and mass from the solar wind environment into Earth’s magnetosphere (see, for instance, Ref. \cite{has} and references therein).\\

  In recent Ref. \cite{nak} results of a 3D kinetic simulation of the Kelvin-Helmholtz instability, with the conditions relevant to the Magnetospheric Multiscale spacecraft (MMS) observations in the Earth's magnetopause \cite{has}, were reported. Figure 29 shows the power spectrum of a magnetic field component computed in the numerical simulation and corresponding to the magnetic field component normal to the nominal magnetopause in the MMS observations. The spectral data were taken from Fig. 4b of the Ref. \cite{has}. The $\lambda_{KH}$ denotes the most unstable Kelvin-Helmholtz instability wavelength, and the ion gyroradius $\rho_i$ and ion inertial length $d_i$ are indicated in Fig. 29 by the vertical dotted lines. The dashed curve indicates the stretched exponential spectrum Eq. (22).\\
 
  In the Ref. \cite{has} results of measurements produced by the Magnetospheric Multiscale spacecraft (MMS)  in the Kelvin-Helmholtz vortices at the magnetopause were reported. Figure 30 shows the corresponding power spectrum of the magnetic field component perpendicular to the mean magnetic field and normal to the nominal magnetopause for the Kelvin-Helmholtz active period. The spectral data were taken from Fig. 3a of the Ref. \cite{has}. The proton gyroradius $\rho_p$ and proton inertial length $d_p$ are indicated in Fig. 30 by the vertical dotted lines. The dashed curve indicates the stretched exponential spectrum Eq. (23).\\
  
  It is interesting that for both the numerical simulation Fig. 29 and the real magnetospheric measurements Fig. 30 the magnetic helicity-dominated spectral laws Eq. (22) and Eq. (23) were observed. For the real measurements, however, a significant role of the background magnetic field is apparent.

 \section{Magnetosphere of other planets: Saturn, Jupiter, and Mercury}

\begin{figure} \vspace{-0.4cm}\centering
\epsfig{width=.45\textwidth,file=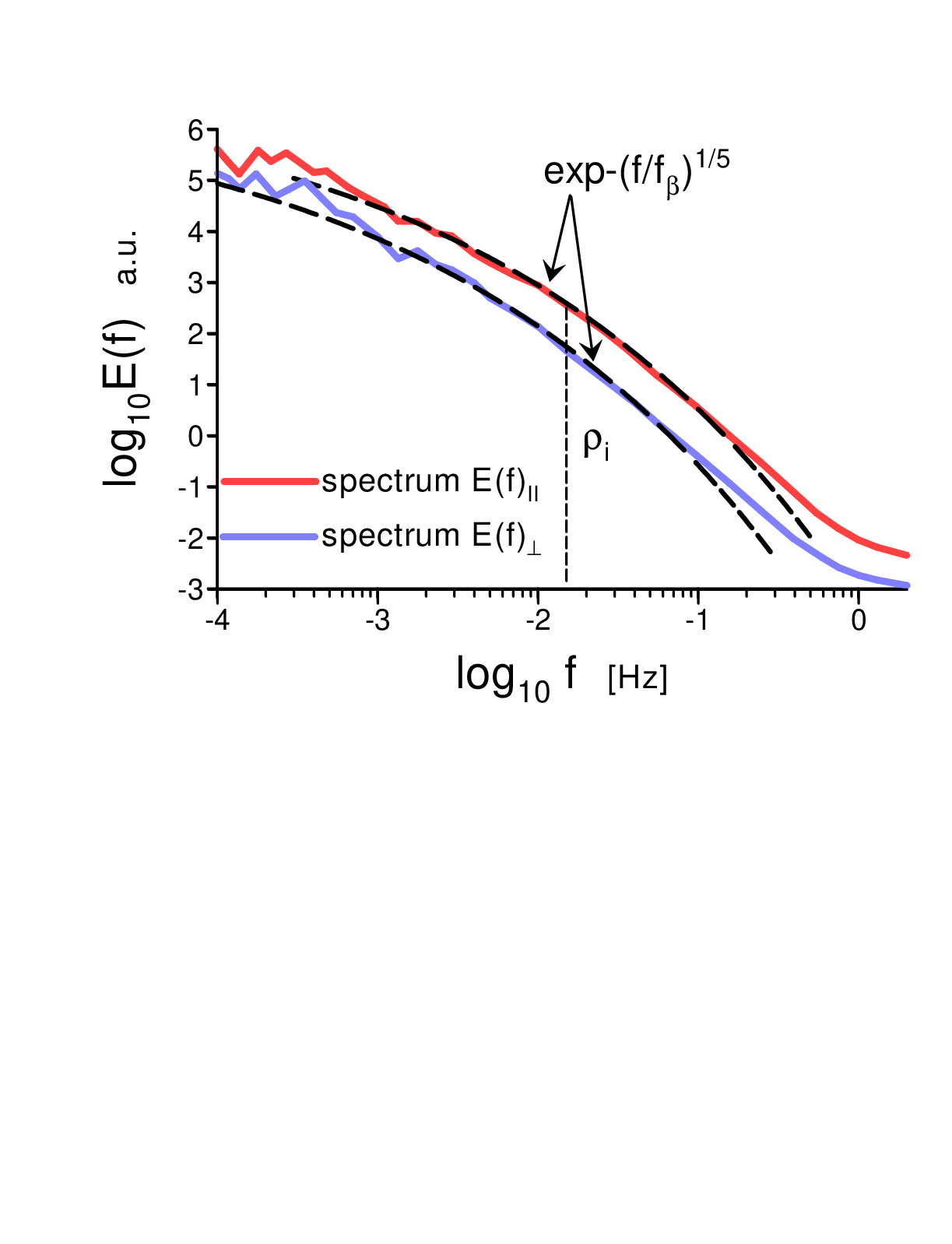} \vspace{-4.6cm}
\caption{Power spectra of the magnetic field fluctuations components corresponding to the planetary ring current $r > 9R_S$ (Saturn, Cassini mission).}
\end{figure}
\begin{figure} \vspace{-0.8cm}\centering
\epsfig{width=.45\textwidth,file=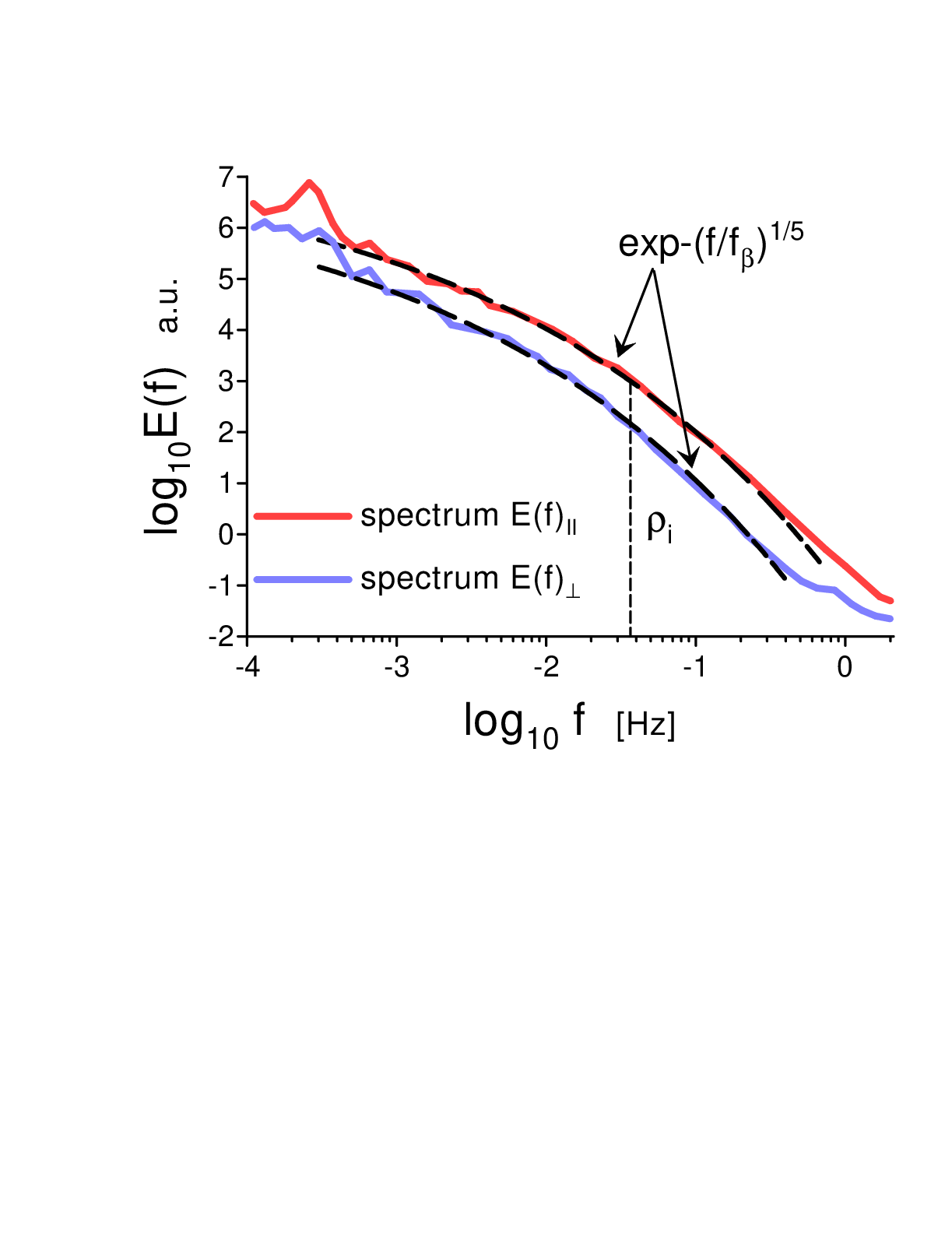} \vspace{-4.2cm}
\caption{Power spectra of the magnetic field fluctuations components corresponding to the plasma dynamics throughout the time of isolated flux tube interchanges (Saturn, Cassini mission).}
\end{figure}
\begin{figure} \vspace{-0.5cm}\centering
\epsfig{width=.45\textwidth,file=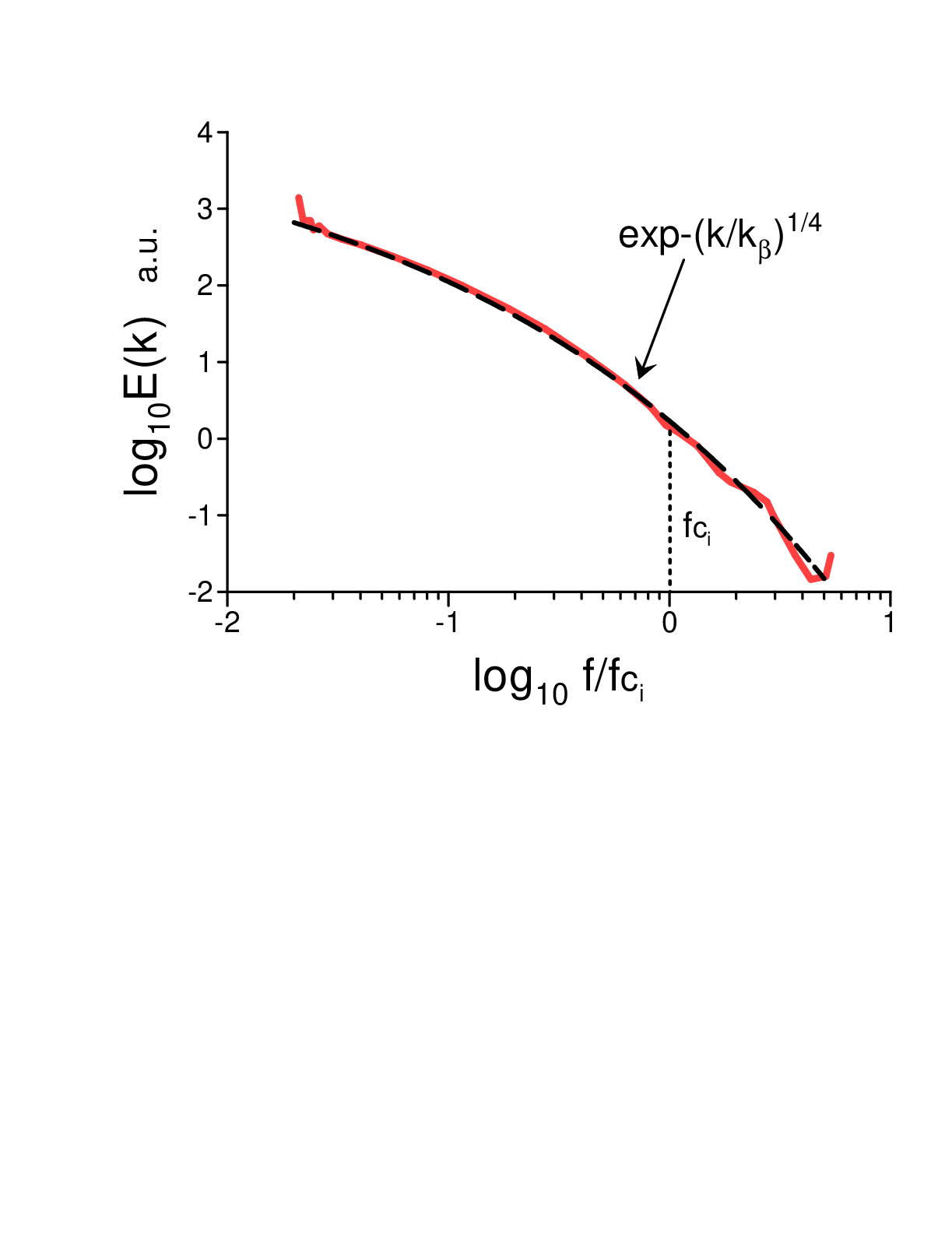} \vspace{-4.6cm}
\caption{Power spectra of the magnetic field fluctuations measured in the magnetosphere of Jupiter at the radial distance bin $20 - 25R_j$ (Galileo mission). }
\end{figure}

  Planets Saturn, Jupiter, and Mercury have their own magnetic field and magnetosphere. The physical conditions in Saturn’s magnetosphere are rather different from those in the solar wind and Earth's magnetosphere. The background magnetic field in this case is strong whereas the magnetic field fluctuations are rather weak. While the Earth's magnetosphere is mainly driven by the surrounding solar wind Saturn’s magnetosphere is mainly driven by the planet's rotation. For the middle magnetosphere $6R_S< r < 15R_S $ (which will be considered in this subsection) a ring current and a flux tube interchange also play a significant role. Due to the fast rotation of the planet, the magnetospheric plasma is mainly concentrated in a plasma sheet located in the equatorial plane. Considering these circumstances it is interesting whether the above-discussed magnetic helicity-determined spectral laws will also take place in Saturn's magnetosphere as well. \\

  In a paper Ref. \cite{psa} results of a statistical analysis of the measurements produced by the Cassini spacecraft in Saturn’s middle magnetosphere (inside the plasma sheet) were reported. Figure 31 shows the power spectra of the magnetic field fluctuations components parallel and perpendicular to the mean magnetic field. The spectral data were taken from Fig. 6a of the Ref. \cite{psa} (the inbound leg of the Cassini spacecraft's second orbit). The data correspond to the planetary ring current $r > 9R_S$. The dashed curves indicate the stretched exponential spectrum Eq. (23). \\
  
    Figure 32 also shows the power spectra of the magnetic field fluctuations components parallel and perpendicular to the mean magnetic field. The spectral data were taken from Fig. 6c of the Ref. \cite{psa}  (the inbound leg of the Cassini spacecraft's second orbit). The data correspond to the plasma dynamics throughout the time of isolated flux tube interchanges. The dashed curves indicate the stretched exponential spectrum Eq. (23). \\
    
    Despite significant differences in the physical conditions of Saturn's magnetosphere from those in the solar wind and Earth's magnetosphere, the magnetic helicity-determined spectral law Eq. (23) is valid there as well.\\
    
    The analogous situation takes also place in Jupiter's magnetosphere as measured by a magnetometer (MAG) onboard the Galileo spacecraft. Figure 33 shows the total power spectrum of the magnetic field fluctuations computed using the magnetometer's data. The spectral data corresponding to the radial distance bin $20 - 25R_j$ ($R_j$ is Jupiter's radius) were taken from Fig. 10b of the Ref. \cite{tao}.  The dashed curve in the Fig. 33 indicates the spectrum Eq. (22). The position of the local ion cyclotron frequency $f{\tiny c}_i$ is indicated by the dotted vertical line. \\

    Figure 34 shows the power spectra of the magnetic field fluctuations computed using the measurements with the magnetometer MAG onboard the Messenger spacecraft in the Mercury's plasma environment. The spectral data were taken from Figs. 6 of the Ref. \cite{huang3}. The dashed curves in the Fig. 34 indicate the spectrum Eq. (23) for the magnetosphere and for the magnetosheath close to bow shock, and spectrum Eq. (22) for the magnetosheath close to magnetopause, respectively. The $\rho_p$ (indicated by the vertical dashed line) corresponds to the proton Larmor radius for the magnetosphere.

\section{Solar active regions}

\begin{figure} \vspace{-0.9cm}\centering
\epsfig{width=.45\textwidth,file=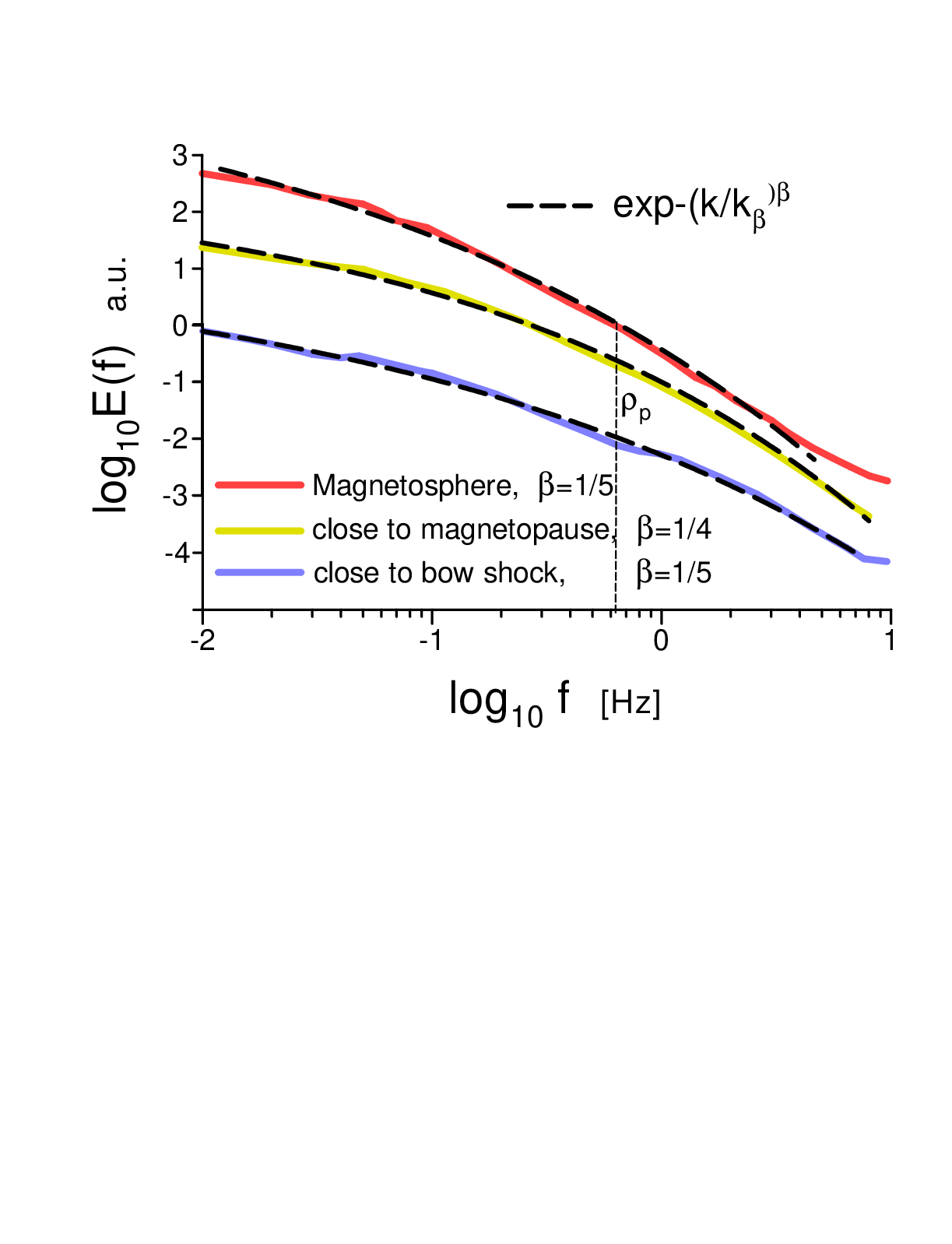} \vspace{-4.45cm}
\caption{Power spectrum of the magnetic field fluctuations measured in the Mercury's plasma environment (Messenger mission). }
\end{figure}
\begin{figure} \vspace{-0.5cm}\centering
\epsfig{width=.45\textwidth,file=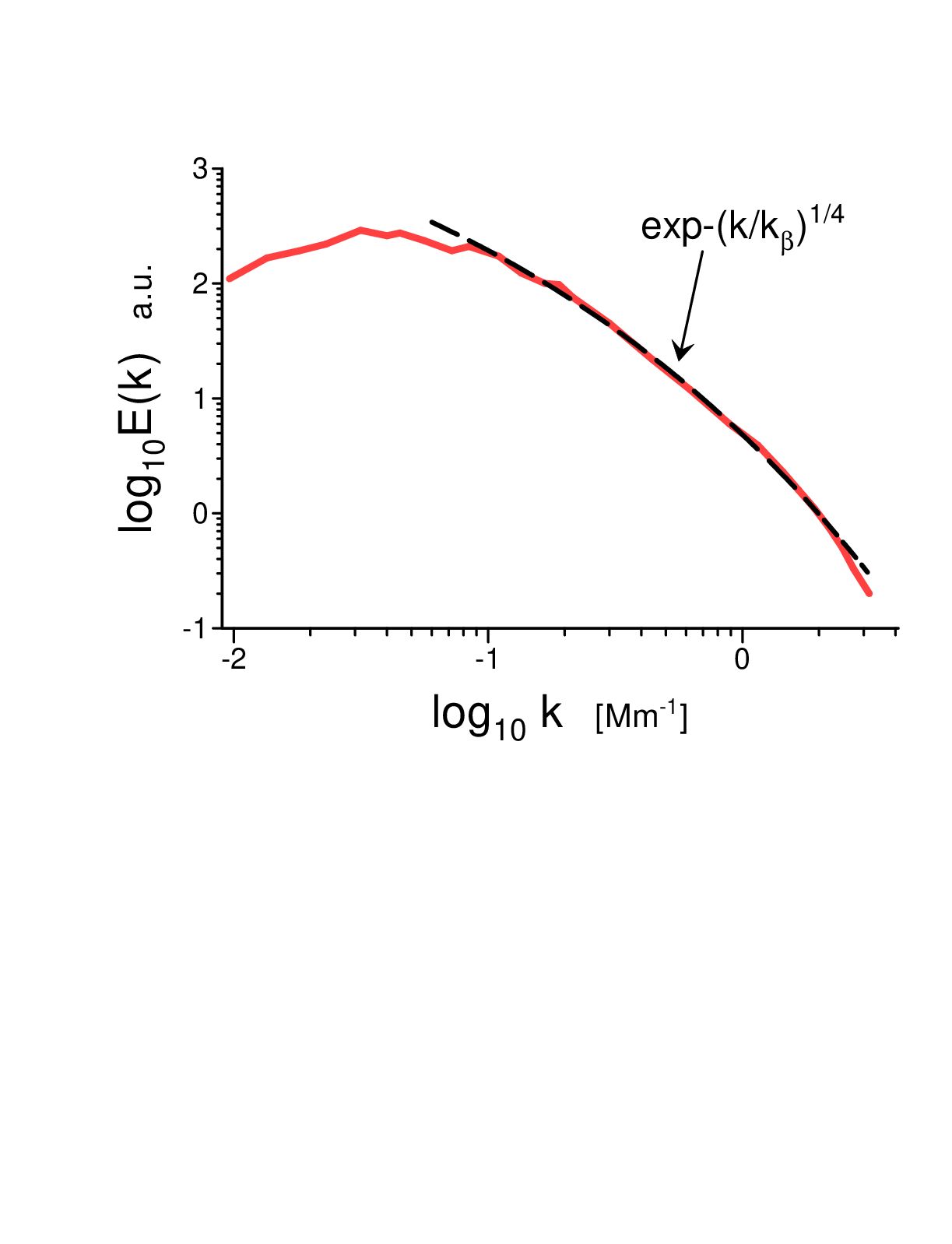} \vspace{-4.32cm}
\caption{Magnetic energy spectrum for synoptic maps of the early declining phase of solar cycle 24 (Helioseismic and Magnetic Imager
onboard the Solar dynamics observatory).}
\end{figure}
\begin{figure} \vspace{-0.5cm}\centering
\epsfig{width=.45\textwidth,file=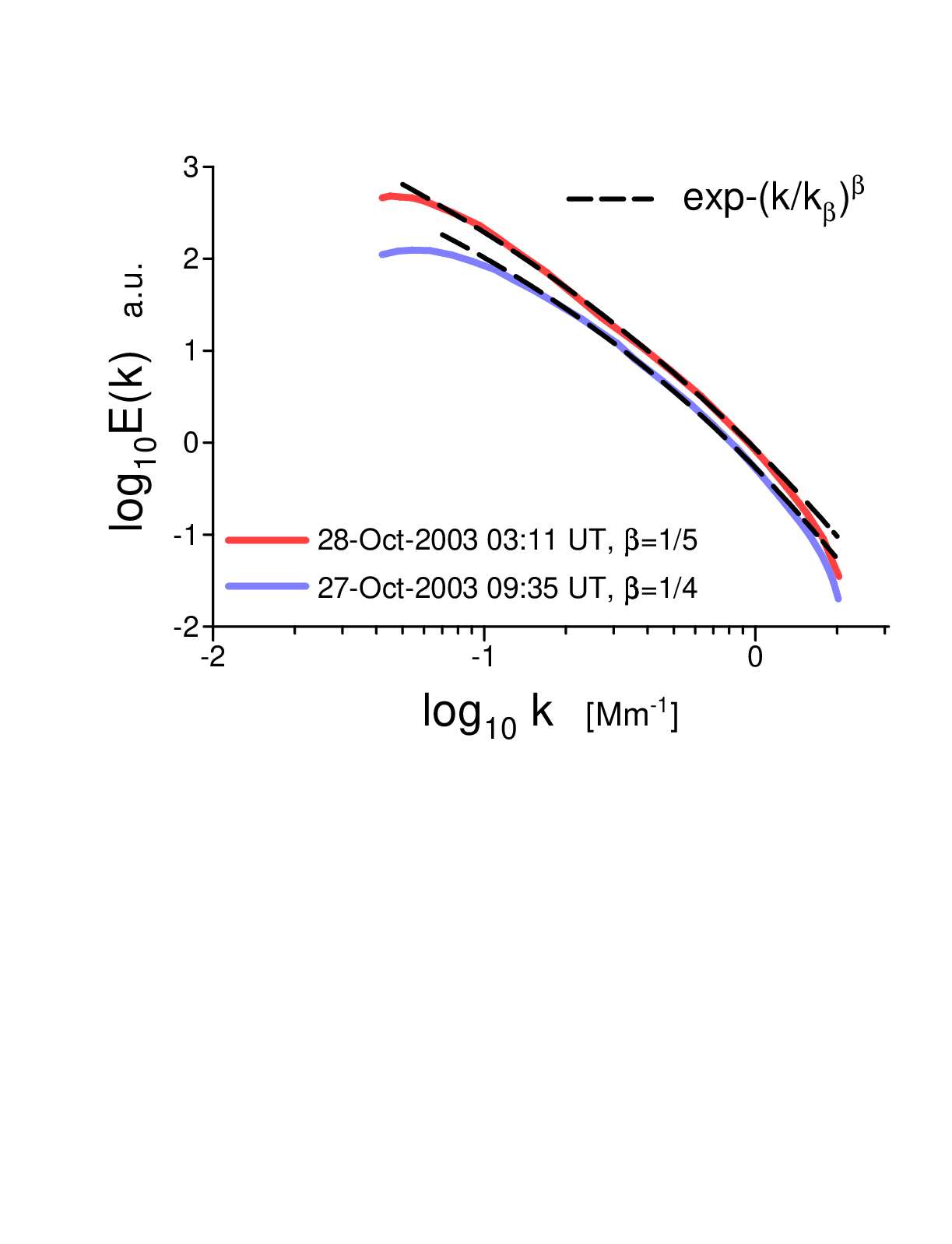} \vspace{-4.35cm}
\caption{Magnetic energy spectra for a rapidly emerging active region NOAA 10488 at a mature stage of its evolution. }
\end{figure}

  The magneto-inertial range of scales can be also observed for the magnetic fields in the mature and intensive solar active regions. \\
  
Figure 35 shows the magnetic energy spectrum for synoptic maps of early declining phases of solar cycle 24. In the time of the early declining phase both magnetic energy and magnetic helicity have their maxima \cite{singh}. The spectral data were taken from Figs. 4.2b of Ref. \cite{prabhu}. The data used for computing the spectrum were obtained by the Helioseismic and Magnetic Imager onboard the Solar dynamics observatory. \\
  
  The dashed curve indicates the best fit of the spectrum with Eq. (22) ($\beta =1/4$, the magneto-inertial range of scales).\\

  Figure 36 shows the magnetic energy spectra for a rapidly emerging active region NOAA 10488 at a mature stage of its evolution. The spectral data for this figure were taken from Fig. 8 of a paper Ref. \cite{hew}. The solar magnetogram images used for the computation of the spectra were obtained by the Michelson Doppler Imager onboard the Solar and Heliospheric Observatory spacecraft. Two C-class flares and a M-class flare occurred in the solar active region, and also a solar active region NOAA 10493 appeared near the active region  NOAA 10488 during 27-28 October 2003.\\ 
  
   The dashed curves indicate the best fit with the Eq. (22) ($\beta =1/4$) and Eq. (23) ($\beta =1/5$), corresponding to the magneto-inertial range of scales at the mature stage (27-28 October 2003) of the active region NOAA 10488. One can see that a mean magnetic field becomes influential on 28 October.\\
   
    Figure 37 shows the magnetic energy spectra at two moments of the emergence of solar active region NOAA 12219. The spectral data were taken from Fig. 1 of a recent paper Ref. \cite{kak}. The solar magnetogram images used for the computation of the spectra were obtained by the Helioseismic and Magnetic Imager onboard the Solar Dynamics Observatory spacecraft. \\

The time $t_0$ corresponds to the onset of the emergence and the time $t_2$ corresponds to the moment when the magnetic flux has its maximum. \\
    
   The spectra in the Fig. 37 are fitted by the spectral laws (the dashed curves): the deterministic chaos Eq. (2) ($\beta =1$), and the magneto-inertial range of scales under the strong influence of a mean magnetic field Eq. (43)  ($\beta =1/5$). \\
   
   Figure 38 shows the evolution of the magnetic energy spectra during the emergence of a large solar active region NOAA 11726. The spectral data for this figure were taken from Fig. 2a of paper Ref. \cite{kka}. The solar magnetogram images used for the computation of the spectra were obtained by Helioseismic and Magnetic Imager onboard of the Solar Dynamics Observatory. \\

  The spectra are fitted by a sequence of the spectral laws (shown as the dashed curves): the time $t_0$ corresponds to the onset of the emergence - the deterministic chaos Eq. (5) ($\beta =1$), later the helical distributed chaos Eq. (12) ($\beta =1/2$), and at a mature stage the magneto-inertial range of scales Eq. (22) ($\beta =1/4$). 
  
 \section{Temporal  variability of the magnetic field at Earth's surface}
 
   Until now the spatial properties of the magneto-inertial range have been studied (the above-discussed frequency spectra reflect these properties due to the Taylor's `frozen-in' hypothesis). It is interesting also to study the true temporal spectral properties of the magneto-inertial range. Temporal variability of the magnetic field at Earth's surface provides such a possibility. In a wide range of time scales this variability is dominated by the variability of the external magnetic field arising from the interaction of the near-Earth solar wind with Earth’s magnetosphere and its inner edge - ionosphere (see, for instance, a recent paper \cite{cc} and references therein).\\

\begin{figure} \vspace{-0.8cm}\centering
\epsfig{width=.45\textwidth,file=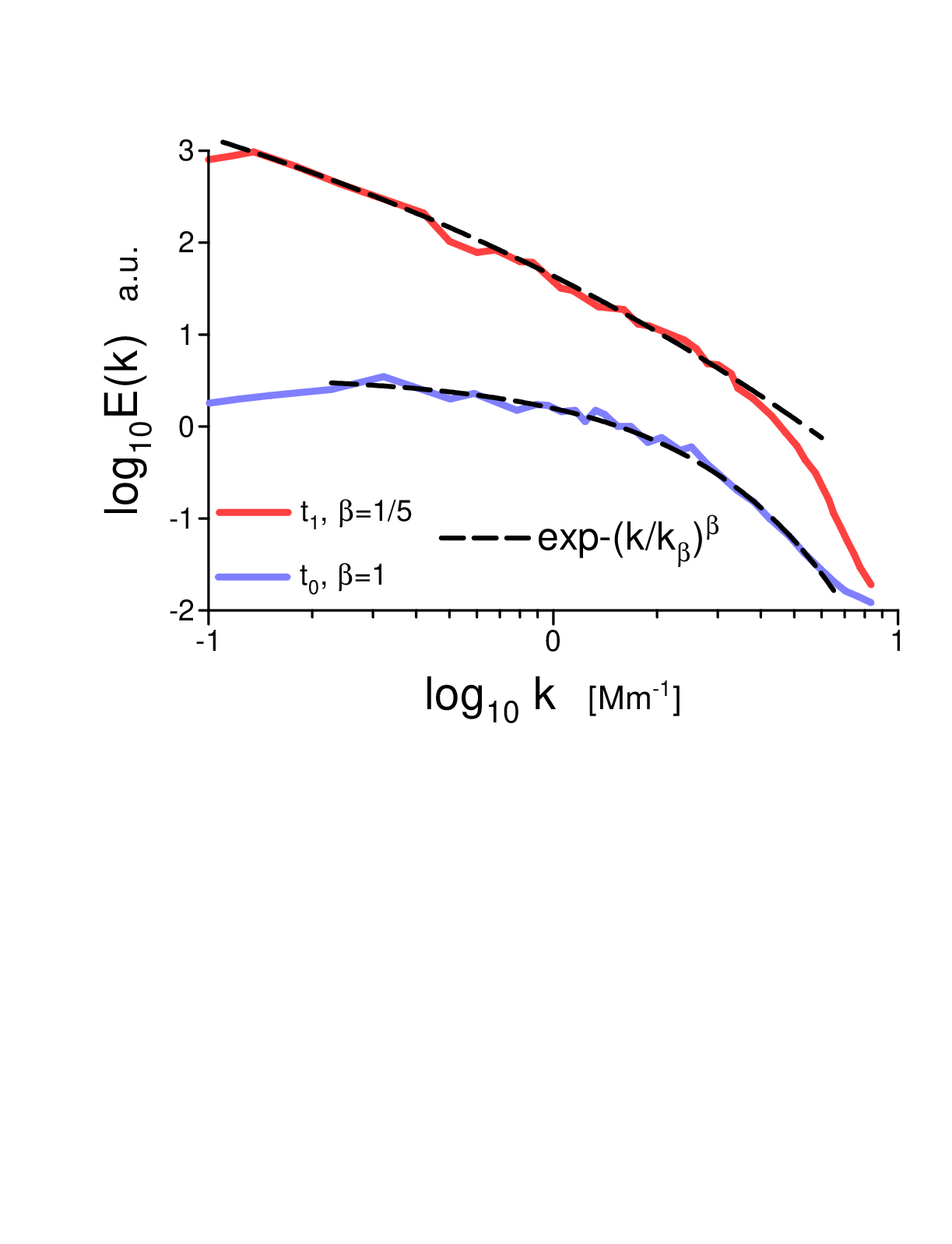} \vspace{-4.5cm}
\caption{Magnetic energy spectra for the emerging active region NOAA 12219:  $t_0$ is the time of the emergence onset (bottom) and $t_1$ is the time of the maximal magnetic flux (top). }
\end{figure}
\begin{figure} \vspace{-0.5cm}\centering
\epsfig{width=.45\textwidth,file=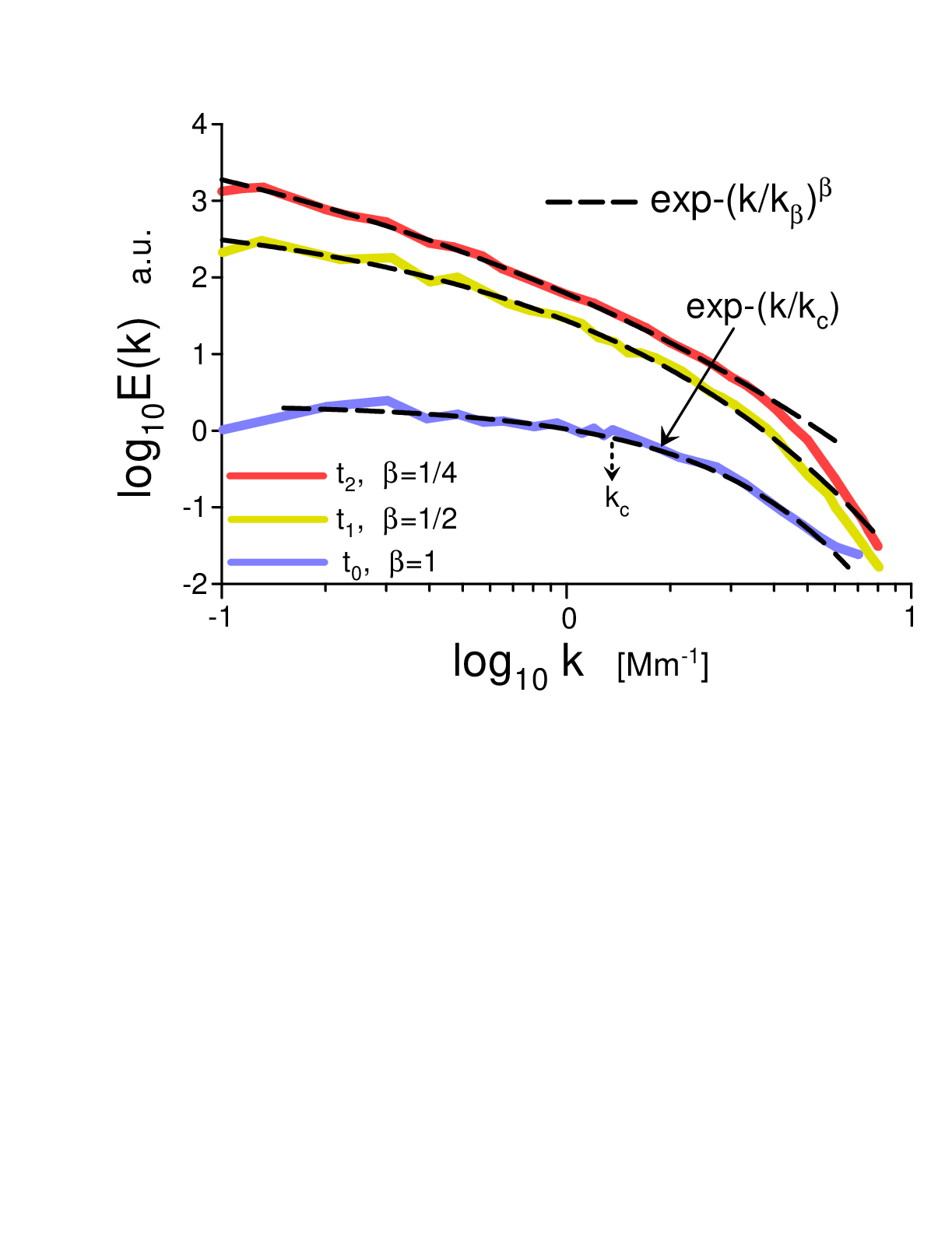} \vspace{-4.75cm}
\caption{Magnetic energy spectra during the emergence of a large solar active region NOAA 11726. }
\end{figure}
   
   To find the true frequency (temporal) spectra corresponding to the magneto-inertial range one should replace the Eqs. (16), (17) and (18) by equations 
\bea
 B_c \propto \varepsilon_{h_m}^{1/2} ~\varepsilon^{-1/4}~f_c^{1/4},  
\eea
\bea
 B_c \propto  \varepsilon_{\hat{h}_m}^{1/2}~ (\varepsilon \widetilde{B}_0)^{-1/6}  f_c^{1/6}, 
\eea
and
\bea
 B_c \propto f_c^{\alpha}   
 \eea
 where $f_c$ is a true characteristic frequency. \\
 
   Hence $\alpha =1/4$ for the case represented by Eq. (32) and $\alpha = 1/6$ for the case represented by Eq. (33). It follows from the Eq. (21) (which is obviously valid for the frequency spectra as well) that $\beta =1/3$ for the case of the Eq. (32) and $\beta =1/4$ for the case of the Eq. (33). Corresponding frequency spectra have the form
\bea
 E(f) \propto \exp-(f/f_{\beta})^{1/3}  
\eea
 and
\bea
 E(f) \propto \exp-(f/f_{\beta})^{1/4}  
\eea  
for the Eq. (32) and Eq. (33) respectively.  \\

   Figure 39 shows the true frequency magnetic energy spectrum obtained using measurements provided by the global magnetic observatory network (the spectral data were taken from Fig. 1 of the paper \cite{cc}). The range of frequencies shown in the Fig. 39 corresponds to the above-mentioned time scales for which variability of the magnetic field   at Earth surface is dominated by the variability of the external magnetic field arising from the interaction of the near-Earth solar wind with Earth’s magnetosphere and ionosphere.  \\
   
   The 27-day peak and its harmonics are related to the solar rotation period, whereas the peak and its harmonics corresponding to daily variations are presumably associated with thermal electric currents in the ionosphere on the heated day side of the globe. Seasonal variations of the ionosphere heating can be the main reason for the appearance of the spectral peaks corresponding to the 1 year and 6 months (see Ref. \cite{cc} and references therein). \\
   
   The dashed curve in the Fig. 39 indicates correspondence to the spectral law Eq. (36) taking into account the influence of a mean magnetic field on the magneto-inertial range of scales.

\section{Conclusions}

  The variability of the spectral laws of the magnetic field fluctuations in the space plasmas is determined by the variability of the physical conditions in the solar, solar wind, and planet magnetosphere plasmas. An additional source of variability is the differences between the large-scale (MHD) and small-scale (kinetic) dynamics. Therefore, the existence of certain universality in the abundant observational, experimental, and DNS data can be useful in understanding the underlying physics.  A significant help in the search of this universality can provide the fundamental ideal invariants such as energy and magnetic helicity. \\
  
  It is shown in the present paper that the magnetic helicity-determined spectral laws can provide a reasonable description for a rather wide class of data directly Eq. (12) or in the frames of the Kolmogorov-Iroshnikov phenomenology Eqs. (22), and (23) based on the notion of distributed chaos. \\ 

This approach allows also the description of the transitional range of scales between MHD and kinetics which makes it applicable for the space plasmas where this transition is omnipresent. In the frames of the distributed chaos notion, the transition seems to be rather smooth (at least in the magnetic energy spectra). \\

\begin{figure} \vspace{-0.9cm}\centering \hspace{-1cm}
\epsfig{width=.5\textwidth,file=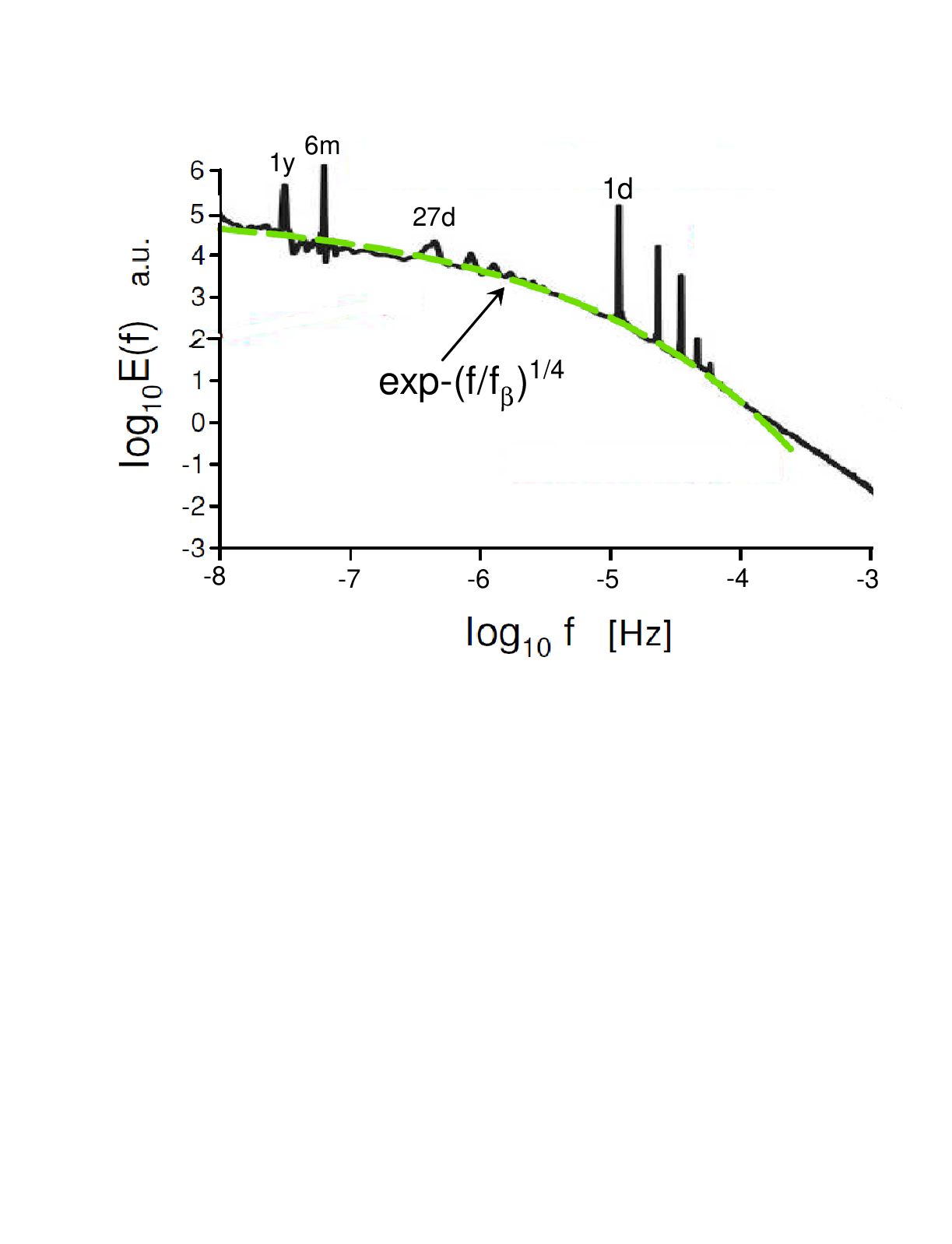} \vspace{-5.7cm}
\caption{The true frequency magnetic energy spectrum obtained using measurements provided by the global observatory network. }
\end{figure}
  Comparing the spectral behavior in the solar wind for different distances from the Sun one can conclude that in the near Sun heliosphere, the role of the background mean magnetic field becomes more prominent Eq. (23) than for the larger distances from the Sun Eq. (22). The transitional distance for this phenomenon seems to be near the value 1.4 AU Fig. 20. \\
  
  In the planets' magnetosphere, the spectral law Eq. (23) is also prevailing, i.e. the role of the background mean magnetic field is generally significant there as well (and also in the intensive solar active regions the role of the background mean magnetic field can become significant at a mature stage of their development). The analogous situation takes also place for the temporal variability of the magnetic field at Earth's surface dominated by the variability of the external magnetic field arising from the interaction of the near-Earth solar wind with Earth’s magnetosphere and ionosphere. \\
  
  Despite the considerable differences in the physical parameters and scales, the results of numerical simulations are in quantitative agreement with the observational data in the frames of the magneto-inertial range notion.

 \newpage

\section{Acknowledgments }

  I thank R. Grappin, H.K. Moffatt, and J.V. Shebalin for stimulating discussions.

\end{document}